\documentclass[prl,reprint,superscriptaddress]{revtex4-1}
\usepackage[pdftex,colorlinks=true,bookmarks=false,citecolor=blue,urlcolor=blue]{hyperref}
\usepackage{pdfpages}
\usepackage{amsmath}
\usepackage{amssymb}
\usepackage{graphicx}

\usepackage{hyperref}
\usepackage{bm}
\usepackage{color}
\usepackage{pifont}

\newcommand{\mysize}{3.4}

\makeatletter
\AtBeginDocument{\let\LS@rot\@undefined}
\makeatother

\begin{document}


\title{Dense Polarized Positrons from Laser-Irradiated Foil Targets in the QED Regime}

\affiliation{Beijing National Laboratory for Condensed Matter Physics, Institute of Physics, Chinese Academy of Sciences, Beijing 100190, China}
\affiliation{Department of Physics and Beijing Key Laboratory of Opto-electronic Functional Materials and Micro–nano Devices, Renmin University of China, Beijing 100872, China}
\affiliation{School of Physical Sciences, University of Chinese Academy of Sciences, Beijing 100049, China}
\affiliation{Collaborative Innovation Center of IFSA, Shanghai Jiao Tong University, Shanghai 200240, China}
\affiliation{Songshan Lake Materials Laboratory, Dongguan, Guangdong 523808, China}

\author{Huai-Hang Song}
\affiliation{Beijing National Laboratory for Condensed Matter Physics, Institute of Physics, Chinese Academy of Sciences, Beijing 100190, China}
\affiliation{School of Physical Sciences, University of Chinese Academy of Sciences, Beijing 100049, China}

\author{Wei-Min Wang} \email{weiminwang1@ruc.edu.cn}
\affiliation{Department of Physics and Beijing Key Laboratory of Opto-electronic Functional Materials and Micro–nano Devices, Renmin University of China, Beijing 100872, China}
\affiliation{Collaborative Innovation Center of IFSA, Shanghai Jiao Tong University, Shanghai 200240, China}

\author{Yu-Tong Li}
\email{ytli@iphy.ac.cn}
\affiliation{Beijing National Laboratory for Condensed Matter Physics, Institute of Physics, Chinese Academy of Sciences, Beijing 100190, China}
\affiliation{School of Physical Sciences, University of Chinese Academy of Sciences, Beijing 100049, China}
\affiliation{Songshan Lake Materials Laboratory, Dongguan, Guangdong 523808, China}

\date{\today}

\begin{abstract}
Dense positrons are shown to be effectively generated from
laser-solid interactions in the strong-field quantum electrodynamics
(QED) regime. Whether these positrons are polarized has not yet been
reported, limiting their potential applications. Here, by QED
particle-in-cell simulations including electron-positron spin and
photon polarization effects, we investigate a typical laser-solid
setup that an ultraintense linearly polarized laser irradiates a foil target with
$\mu$m-scale-length preplasma. We find that once the positron yield
becomes appreciable with the laser intensity exceeding $10^{24}~\rm
W/\rm cm^2$, the positrons are obviously polarized. The polarized positrons
can acquire $>30\%$ polarization degree and $>30$ nC charge with a flux of
$10^{12}\,{\rm sr}^{-1}$. The polarization relies on the deflected
angles and can reach 60\% at some angles and energies. The angularly-dependent
polarization is attributed to the asymmetrical laser fields
positrons undergo in the skin layer of overdense plasma, where the
radiative spin-flip and radiation reaction play significant roles.
The positron polarization is robust and could generally appear in
future 100-PW-class laser-solid experiments for various
applications.
\end{abstract}


\maketitle

\newpage

Polarized positrons with a preferential orientation of spins can
exhibit unique features in many areas, such as searching new physics
beyond the Standard Model in International Linear
Collider (ILC) \cite{Moortgat2008pr, Vauth2016ijmp} and probing spin phenomena at material surfaces
\cite{Gidley1982prl,Hugenschmidt2016ssr}. Besides, polarized
electron-positron ($e^-e^+$) plasmas are believed to be
ubiquitous in pulsar magnetospheres \cite{Novak2009prd}. There are a
few methods to generate high-energy polarized positrons.
Ultrarelativistic positrons in tesla-level magnetic fields of
storage rings can be polarized via radiative spin-flip
\cite{Sokolov1968qed,Baier1972spu} but rather slowly. Alternatively, polarized positrons are usually produced via Bethe-Heitler
(BH) process by hitting high-Z targets with circularly polarized
$\gamma$ photons \cite{Alexander2008prl,Omori2006prl} or
prepolarized electrons \cite{Abbott2016prl}. These BH methods suffer
low conversion efficiency of $\sim10^4$ positrons ($10^{-6}$ nC)
per shot, and thus high repetitions are necessary to meet the
high-charge or -density requirements of ILC (3.2 nC) and
laboratory astrophysics.

Dense positrons can be efficiently generated from single-shot
laser-matter interactions in the strong-field quantum
electrodynamics (QED) regime
\cite{Berestetskii1982qed,Baier1998qed,Ritus1985jslr,piazza2012rmp,Gonoskov2021arxiv}.
This approach is becoming experimentally feasible with advances in
high-intensity laser technologies \cite{Danson2019hplse}. Recently, the intensity of $1.1\times10^{23}~\rm W/\rm cm^2$ has
been realized by a 4-PW laser system \cite{Yoon2021optica}, and higher-power
laser facilities of 10-PW \cite{ELI} to 100-PW classes will be available
\cite{Cartlidge2018science,Shao2020ol,XCELS,Cartlidge2017science}.
In such strong laser fields, $\gamma$ photons can be radiated by electrons and in turn annihilate into $e^-e^+$ pairs via Breit-Wheeler (BW) process \cite{Reiss1962jmp}. For all-optical
configurations of lasers colliding with unprepolarized multi-GeV
electrons \cite{Sokolov2010prl,Blackburn2017pra}, polarized
positrons can be obtained if asymmetric laser fields are employed,
such as elliptically polarized \cite{Wan2020plb} or two-color
linearly polarized laser pulses \cite{Chen2019prl}. Limited by the
charge of electron beams from laser wakefield accelerators, the
corresponding positron yield is at the $10^{-4}$ nC level.
Furthermore, constructing such asymmetric strong laser fields is
challenging due to the low damage threshold of optical devices
\cite{Tien1999prl}. Recent QED particle-in-cell (PIC) simulations
have shown that impinging on a stationary target by two
counter-propagating 10-PW-class lasers
\cite{Nerush2011prl,Zhu2016nc,Grismayer2017pre} or one 100-PW-class
laser \cite{Ridgers2012prl,Kostyukov2016pop,Wang2017pre} is capable
of generating much denser positrons over 100 nC via self-sustained
QED cascades. However, it remains unclear whether such positrons are
polarized or not because the QED model being widely adopted in the
existing QED-PIC codes
\cite{Elkina2011prab,Ridgers2014jcp,Gonoskov2015pre,Wang2017pre}
overlooks the positron spin dynamics.

\begin{figure}[t]
\centering
\includegraphics[width=\mysize in]{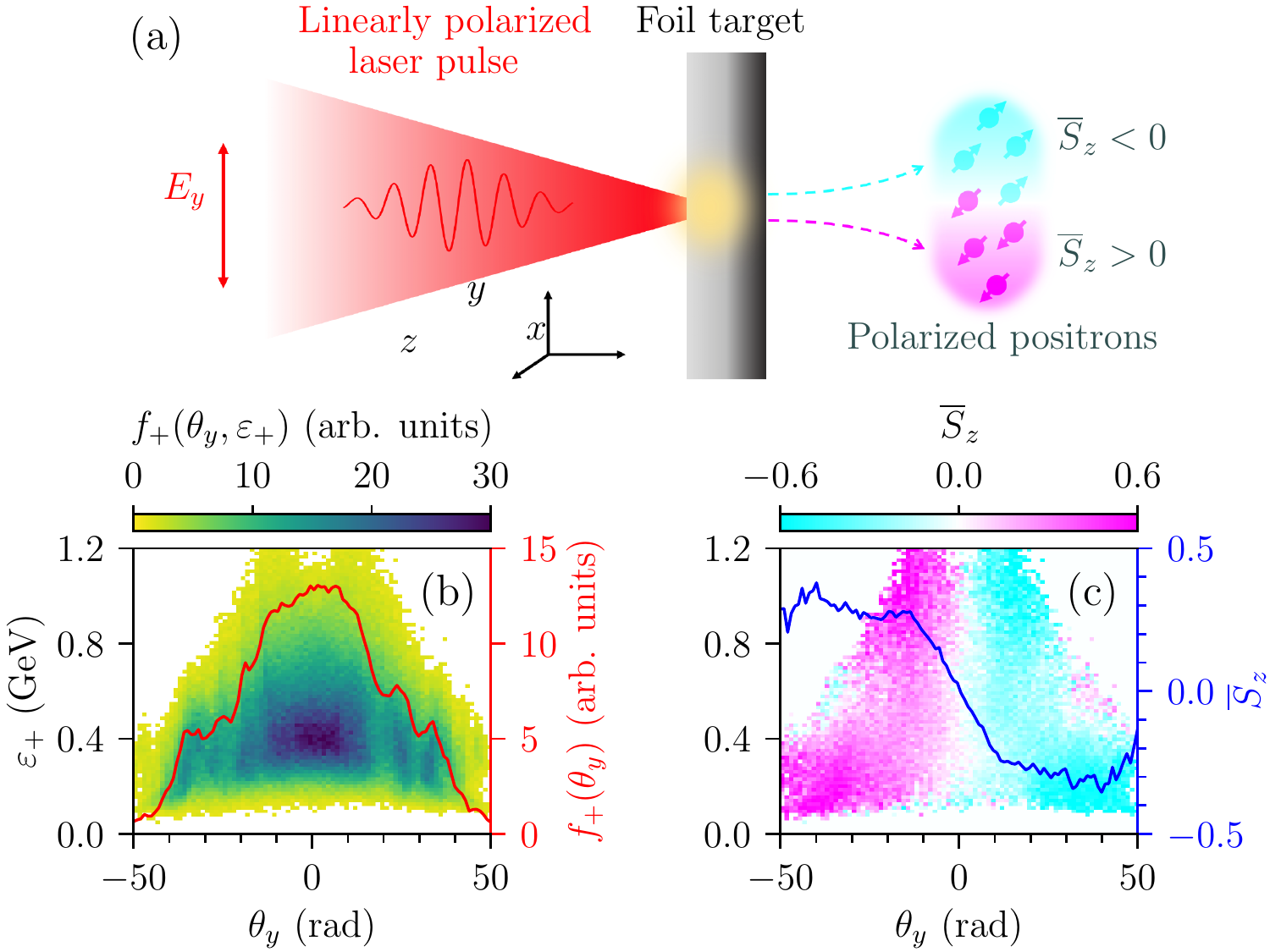}
\caption{\label{fig1}(a) Schematic for generating polarized positrons in laser-solid interactions, where a linearly polarized laser pulse impinges on a foil target
with $\mu$m-scale-length preplasma. Positrons of two opposite
polarizations $\overline S_z>0$ and $\overline S_z<0$ are generated
at opposite deflection angles $\theta_y<0$ and $\theta_y>0$,
respectively, where $\theta_y=\arctan(p_y/|p_x|)$. (b) and (c) are
the positron number distribution $f_+(\theta_y,\varepsilon_+)$ and
corresponding polarization distribution $\overline S_z$ versus
$\theta_y$ and energy $\varepsilon_+$ at the end of the interaction, where $f_+(\theta_y)$ and
$\overline S_z$ integrated over $\varepsilon_+$ are plotted by curves.}
\end{figure}

In this Letter, we employ a recently-developed QED-PIC code
including pair spin and photon polarization effects to clarify the
above problem. By QED-PIC simulations, we investigate a linearly
polarized laser interaction with a solid foil target with
$\mu$m-scale-length preplasma, as depicted in Fig.~\ref{fig1}(a).
When the laser intensity exceeds $10^{24}~\rm
W/\rm cm^2$, substantial
positrons are created primarily in the skin layer of solid-density
plasma, where the dominance of laser magnetic components is favorable for $e^-e^+$ pair creation. The positrons are then quickly pushed into
deeper plasma and escape from the laser fields, causing them only
experience subcycle laser fields. In such asymmetric
fields, the created positrons are split into two populations of
opposite spin polarization at the positive and negative deflected
angles, respectively, due to radiative spin-flip and radiation
reaction. Above 30\% polarization of a 30 nC positrons can be
achieved and the polarization can even reach 60\% at some angles and energies.
Note that in future 100-PW laser-solid experiments even aiming at
other applications \cite{Ridgers2012prl,Wang2017pre,Macchi2013rmp}, a skin layer can be
certainly formed, where polarized positron generation could be
ubiquitous, therefore, pair spin and photon polarization effects
should be considered.

\textit{Simulation setups.}---We perform QED-PIC simulations to
investigate the positron polarization sketched above with the code
{\scshape yunic} \cite{Song2021YUNIC,Song2021njp}. Multiphoton Compton scattering and multiphoton BW process can
be characterized by quantum invariant parameters
$\chi_e=(|e|\hbar/m_e^3c^4)|F_{\mu\nu}p^{\nu}|$ and
$\chi_\gamma=(|e|\hbar^2/m_e^3c^4)|F_{\mu\nu}k^{\nu}|$,
respectively, where $F_{\mu\nu}$ is the field tensor, $p^{\nu}$
($\hbar k^{\nu}$) is the pair (photon) four-momentum, $\hbar$ is the
reduced Planck constant, $c$ is the speed of light, and $e$ and
$m_e$ are the electron charge and mass. These two leading QED process are implemented through the standard
Monte-Carlo algorithm
\cite{Elkina2011prab,Ridgers2014jcp,Gonoskov2015pre}, but including
$e^-e^+$ spin and photon polarization effects
\cite{Baier1998qed,Li2020prl2,Li2020prl,Cain}. Since we select the
mean axis as quantization axis, the spin vector $\bm S$ of
non-radiating electrons/positrons and Stokes parameter $\bm \xi$ of
non-decaying photons also need to be updated \cite{Cain}. More
implementation details can be found from the Supplemental Material \cite{material}.

We first adopt one-dimensional (1D) PIC simulations to better get
insight into the positron polarization mechanism with higher
numerical resolutions, where the geometry is 1D while $\bm S$, $\bm
\xi$ and the particle momentum $\bm p$ still remain fully
three-dimensional (3D). A laser pulse linearly polarized along the
$y$ direction is normally launched from the left boundary $x=0$ at
the initial time $t=0$. The pulse has a duration $\tau_0=20$ fs
(FWHM of Gaussian temporal envelope), a central wavelength
$\lambda_0=1~\mu$m, and a normalized amplitude
$a_0=eE_L/m_ec\omega_0=1500$ (peak intensity $3\times10^{24}~\rm W/\rm cm^2$), where $E_L$ and $\omega_0$ are the
laser amplitude and frequency. An initially unpolarized and
preionized carbon foil target of an electron density $n_0=530n_c$ is
placed at $x=9.75\lambda_0$ with a thickness $d=0.5~\mu$m, where the critical density
$n_c=m_e\omega_0^2/2\pi e^2\approx1.1\times10^{21}~\rm cm^{-3}$. The results are almost the same with a thicker target since it can not be penetrated by the laser. In the front of the foil target, there is a preplasma following an
exponential density profile with a scale-length $L=1.5~\mu$m. The
simulation domain $L_x=20\lambda_0$, cell size $\Delta
x=\lambda_0/96$, 500 electrons/C$^{6+}$ ions per cell are taken.


\textit{Positron properties.}---By the laser direct acceleration,
some electrons from the preplasma quickly gain hundreds of MeV
energies to emit $\gamma$ photons. Here, the boosted field strength
in the electron rest frame can exceed the Schwinger limit to achieve
$\chi_e\gtrsim1$ and thus it enters the QED-dominated regime. By the
end of the interaction $t=28T_0$, up to $63\%$ of the total laser
energy is transferred to $\gamma$ photons and 18\% to $e^-e^+$
pairs. If the photon polarization and pair spin is not considered,
the pair yield is enhanced by about 10\%, close to our recent PIC
results with counter-propagating laser pulses \cite{Song2021njp}.

Figures \ref{fig1}(b) and \ref{fig1}(c) illustrate that transversely
polarized positrons are obtained and their polarization is angularly
dependent. The positrons deflected along the $\pm y$ directions are
polarized along the $\mp z$ directions, respectively. With
larger deflection angles $|\theta_y|$, positrons generally possess
higher polarization degrees $|\overline S_z|$, which can reach 30\%
for $|\theta_y|>20^\circ$ [the blue line in
Fig.~\ref{fig1}(c)]. More than 50\% of the total positrons acquire a
30\% polarization through the angular selection. In addition,
$|\overline S_z|$ also depends on energy, with higher values in both
lower- and higher-energy regions; it can reach $60\%$ at
some angles and energies, accounting for 1\%
positrons.

\begin{figure}[t]
\centering
\includegraphics[width=\mysize in]{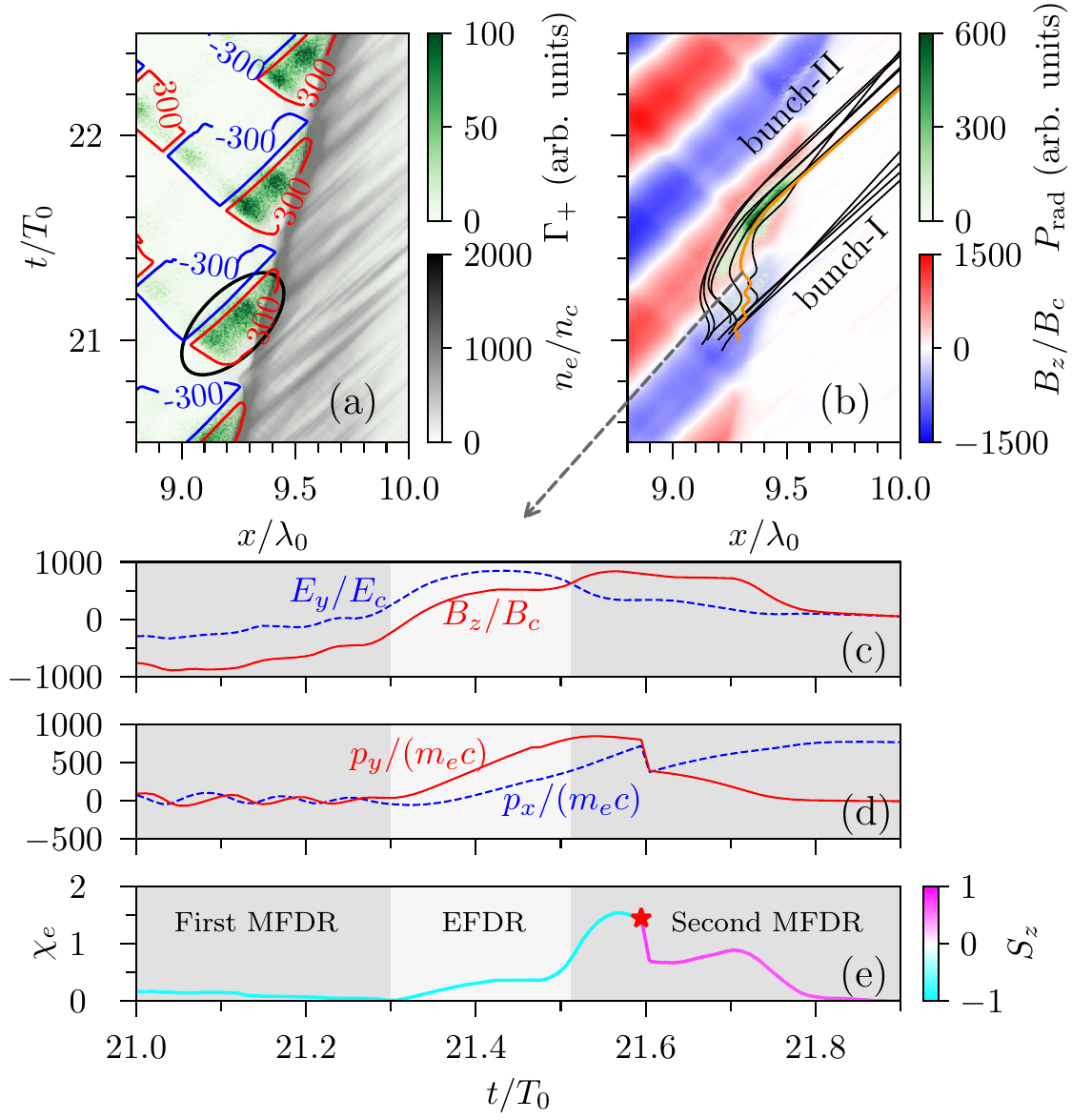}\caption{\label{fig2}
(a) The spatiotemporal evolution of electron density $n_e$ and
positron creation rate $\Gamma_+$, where $\Gamma_+$ represents the
positron number created per unit length and unit time. Contour lines
of $|B_z/B_c|-|E_y/E_c|=\pm300$ are plotted to distinguish EFDR and
MFDR, where $E_c=B_c=m_ec\omega_0/e$. Plots (b)-(e) focus on the
positrons born in the marked elliptical zone in (a). (b) Radiation power $P_{\rm rad}$ of the
marked positrons over the laser magnetic field $B_z$. Eleven positrons are tracked and they can be classified into two bunches,
marked as bunch-I and bunch-II. Evolution of a typical bunch-II
positron [the orange trajectory in (b)]: (c) the experienced fields
$E_y$ and $B_z$, (d) momenta $p_x$ and $p_y$, and (e) QED parameter
$\chi_e$ and spin component $S_z$, where light- and
dark-gray regions denote EFDR and MFDR, respectively, and the
red star in (e) represents a strong radiation event.}
\end{figure}

\textit{Polarization mechanism.}---Positrons are mainly created
in the plasma skin layer near the target front surface, where the
magnetic fields are dominated over the electric ones [Fig.~\ref{fig2}(a)].
The magnetic-field-dominated regime (MFDR)
favors QED processes, while the electric-field-dominated regime
(EFDR) facilitates the $e^-e^+$ acceleration
\cite{Esirkepov2015pla}. Because positrons are discretely created
with a period of half-laser-cycle, we will focus on the
positrons born in the period around $t=21T_0$, as marked by the
elliptical zone in Fig.~\ref{fig2}(a). The obtained results can be
analogously extended to other periods.

\begin{figure}[t]
\centering
\includegraphics[width=\mysize in]{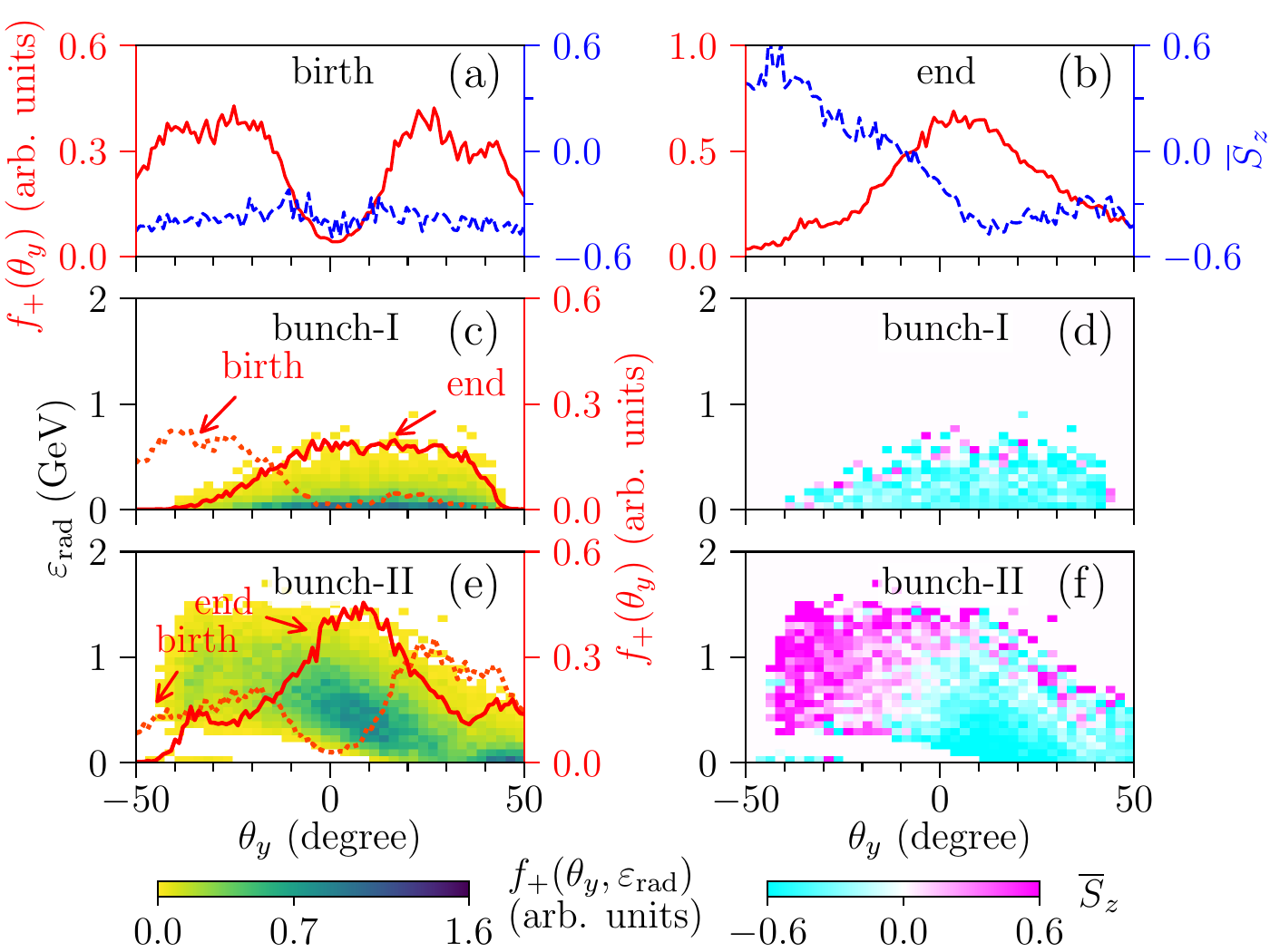}
\caption{\label{fig3}Positrons born in the elliptical zone marked in
Fig.~\ref{fig2}(a) are focused here. The angular distribution of
number $f_+(\theta_y)$ and polarization $\overline S_z$ versus the
deflection angle $\theta_y$ (a) at birth, and (b) the end of the interaction. [(c), (e)]
$f_+(\theta_y,\varepsilon_{\rm rad})$ and [(d),(f)] $\overline S_z$
versus $\theta_y$ and the total radiation energy per positron
$\varepsilon_{\rm rad}$ at the end of the interaction, where [(c), (d)] correspond to bunch-I and [(e), (f)] to bunch-II. Angular
distributions $f_+(\theta_y)$ at birth (red-dotted) and the end of
the interaction (red-solid) are also plotted in (c) for bunch-I and
(e) for bunch-II.}
\end{figure}

In the marked zone, positrons are created in a negative half-cycle
of $B_z<0$, causing their polarization to acquire a negative value
$\overline S_z\approx -0.4$ [Fig.~\ref{fig3}(a)]. This is
because positron spins at birth have higher probabilities to be
parallel to the magnetic field direction $\bm\zeta\equiv{\bm
B}'/|{\bm B}'|$ in their respective rest frames as their parent
$\gamma$ photons of high energies are weakly linearly polarized in
the $x$-$y$ plane \cite{material}. Here, ${\bm B}'=[{\bm E}+\bm \beta
\times \bm B-\bm \beta (\bm \beta \cdot \bm E)] \times {\bm \beta}$
and ${\bm \beta}$ denotes unit vector along the positron
velocity. Considering ${\bm \beta}$ is directed in the $x$-$y$
plane, one can obtain ${\bm\zeta} \approx (0,0,{B_z}/|{B_z}|)$ in
the MFDR, hence $\overline S_z$ of newly created positrons has the
same sign with $B_z$. From Fig.~\ref{fig3}(a), both positron number
and polarization at birth are essentially
symmetrical with respect to $\theta_y=0$. Similarly, positrons born
in adjacent positive half-cycles also exhibit similar distributions,
but with $\overline S_z>0$ due to $B_z>0$ (see Fig.~S5 in
\cite{material}). Thus, the polarization of positrons born at positive
and negative half-cycles could be counteracted each other
and the angularly-dependent polarization would not be achieved if their
spins or deflection angles were not changed later.

The positron polarization in Fig.~\ref{fig1}(c)
is attributed to the asymmetric laser fields that positrons
experience later, where radiative spin-flip and radiation reaction
mainly in the second MFDR play significant roles. The marked
positrons are pushed forward after being created, gradually divide
into bunch-I and bunch-II [Fig.~\ref{fig2}(b)]. Two
bunches successively escape the laser fields from two adjacent
half-cycles, with the relative number $N_+^{\rm I}:N_+^{\rm
II}\approx 3:5$. Only experiencing the first MFDR
where they are born, bunch-I positrons are quickly pushed forward into the deeper
plasma, because their initial negative momenta $p_y<0$ [see the
red-dotted line in Fig.~\ref{fig3}(c)] lead to strong forward
Lorentz forces, i.e., $\beta_yB_z$ along the $+x$ direction. Without
undergoing an EFDR for acceleration, bunch-I is generally less
energetic and weakly radiating [Fig.~\ref{fig3}(c)], hence almost
retaining the initial negative polarization [Fig.~\ref{fig3}(d)]. By
contrast, bunch-II positrons travel through an EFDR to obtain higher
energies and then through the second MFDR to radiate strongly [Fig.~\ref{fig3}(e) and also $P_{\rm rad}$ in Fig.~\ref{fig2}(b)].
Due to quantum stochasticity, spins of only a fraction of bunch-II
positrons can flip parallel to $\bm\zeta$ and achieve an opposite
polarization of $\overline S_z>0$ [Figs.~\ref{fig3}(f)] as
$B_z>0$ in the second MFDR.

Figures~\ref{fig3}(e) and \ref{fig3}(f) show that part bunch-II
positrons with the final $\overline S_z>0$ mainly appear at
$\theta_y<0$, because they undergo strong radiation reaction. This can be
explained by tracking a typical bunch-II positron
[Figs.~\ref{fig2}(c)-\ref{fig2}(e)]. Under the Lorentz force and
radiation reaction, its transverse momentum $p_y$ can be
approximated as $p_y\approx p_{y0}+\int dt\left[|e|(E_y-\beta_xB_z)-\frac{p_y}{\gamma m_ec^2}P_{\rm rad}\right]$, where $p_{y0}$ is the initial $y$-momentum, $\gamma$ is the
relativistic factor, and the last term is radiation reaction
whose direction is opposite to the velocity. Figure~\ref{fig2}(d)
shows that the tracked positron first undergoes the gyration motion
in the first MFDR due to its low initial energy and then enters the
EFDR for a significant acceleration. As $E_y>B_z$ and small
$P_{\rm rad}$ in these two regimes, the positron gains $p_y>0$. After
entering the second MFDR, the positron emits a high-energy photon and simultaneously its $p_y$
is sharply decreased by radiation reaction (i.e., $P_{\rm rad}$ is large enough). As $E_y<B_z$ and
$\beta_x>0$ in the second MFDR, $p_y$ gradually decreases and
changes the sign to achieve $\theta_y<0$. Here, the strong radiation of $\varepsilon_{\rm rad}>0.2~$GeV
is necessary for bunch-II to flip their spins
\cite{Song2019pra} and sharply decrease $p_y$. This
is supported by statistical results in Fig.~\ref{fig3}(e)
that $\theta_y$ tends to change from positive to negative values as
$\varepsilon_{\rm rad}$ increases. Bunch-I finally obtains a small positive deflection angle of
$\overline\theta_y\approx 10^\circ$ in average, making less contribution to the overall polarization.

Therefore, positrons born in negative half-cycles are generally
polarized with $\overline S_z>0$ at $\theta_y<0$ and $\overline
S_z<0$ at $\theta_y>0$ [Fig.~\ref{fig3}(b)] due to the joint action of spin-flip and
radiation reaction. It
also holds for positrons born in positive half-cycles [Fig.~S5(b)
in \cite{material}], hence leading to the overall polarization as
displayed in Fig.~\ref{fig1}(c). Note that the obtained electrons
are also polarized like positrons (Fig.~S4 in \cite{material}),
but with weaker polarizations due to a mixing of unpolarized
target electrons.

\begin{figure}[t]
\centering
\includegraphics[width=\mysize in]{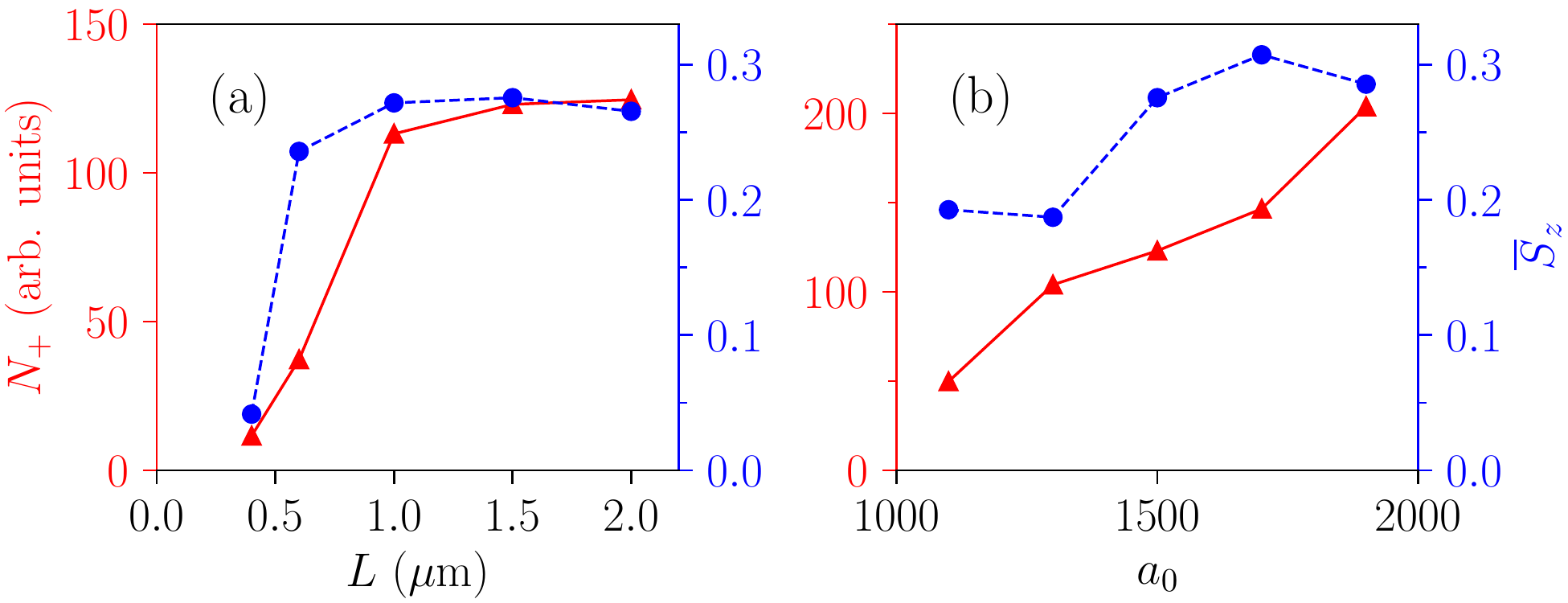}
\caption{\label{fig4}The positron yield $N_+$ (solid-triangle) and
polarization $\overline S_z$ (dashed-circle) at $\theta_y>10^\circ$
versus (a) the preplasma scale-length $L$ and (b) laser amplitude
$a_0$, respectively, where we take $a_0=1500$ in (a) and
$L=1.5~\mu$m in (b). Other parameters are the same as
Fig.~\ref{fig1}.}
\end{figure}

\textit{Parameter influences.}---The dependence of positron yield
$N_+$ and positron polarization $\overline S_z$ on the preplasma
scale-length $L$ and laser amplitude $a_0$ are presented in
Figs.~\ref{fig4}(a) and \ref{fig4}(b), respectively. Preplasmas due to prepulses are unavoidable and also
adjustable in real experiments. Preplasmas of relatively low
densities favor both $N_+$ and $\overline S_z$ by enhancing the
laser absorption and generating more ultrarelativistic electrons to
trigger QED cascades \cite{Wang2017pre}. A small scale-length
preplasma of $L=0.4~\mu$m leads to significant laser reflection, where positrons are mainly created in the strong standing wave away from the target
surface (Fig.~S6 in \cite{material}). It implies that they would
experience quasi-symmetrically multicycle laser fields, which is
detrimental to their polarization. An excessive preplasma of
$L>1.5~\mu$m also causes a slight decrease of $\overline S_z$
[Fig.~\ref{fig4}(a)] since more laser energies are depleted in the
preplasma before reaching the foil target surface.

Figure~\ref{fig4}(b) shows that positrons are polarized once the positron yield becomes appreciable with the laser intensity
above $10^{24}~$W cm$^{-2}$ ($a_0>1000$). $N_+$ increases with the growth of $a_0$ and $\overline S_z$ also tends to
rise until a peak around $a_0=1700$. A similar trend is also
obtained as the target density is changed [see Fig.~S7(b) in
\cite{material}], but the optimized $a_0$ for peaked $\overline S_z$
decreases for a lower density. This is because the laser amplitude
should match with the density to optimize the laser energy
transported in a skin layer, which determines $\overline S_z$.

\begin{figure}[t]
\centering
\includegraphics[width=\mysize in]{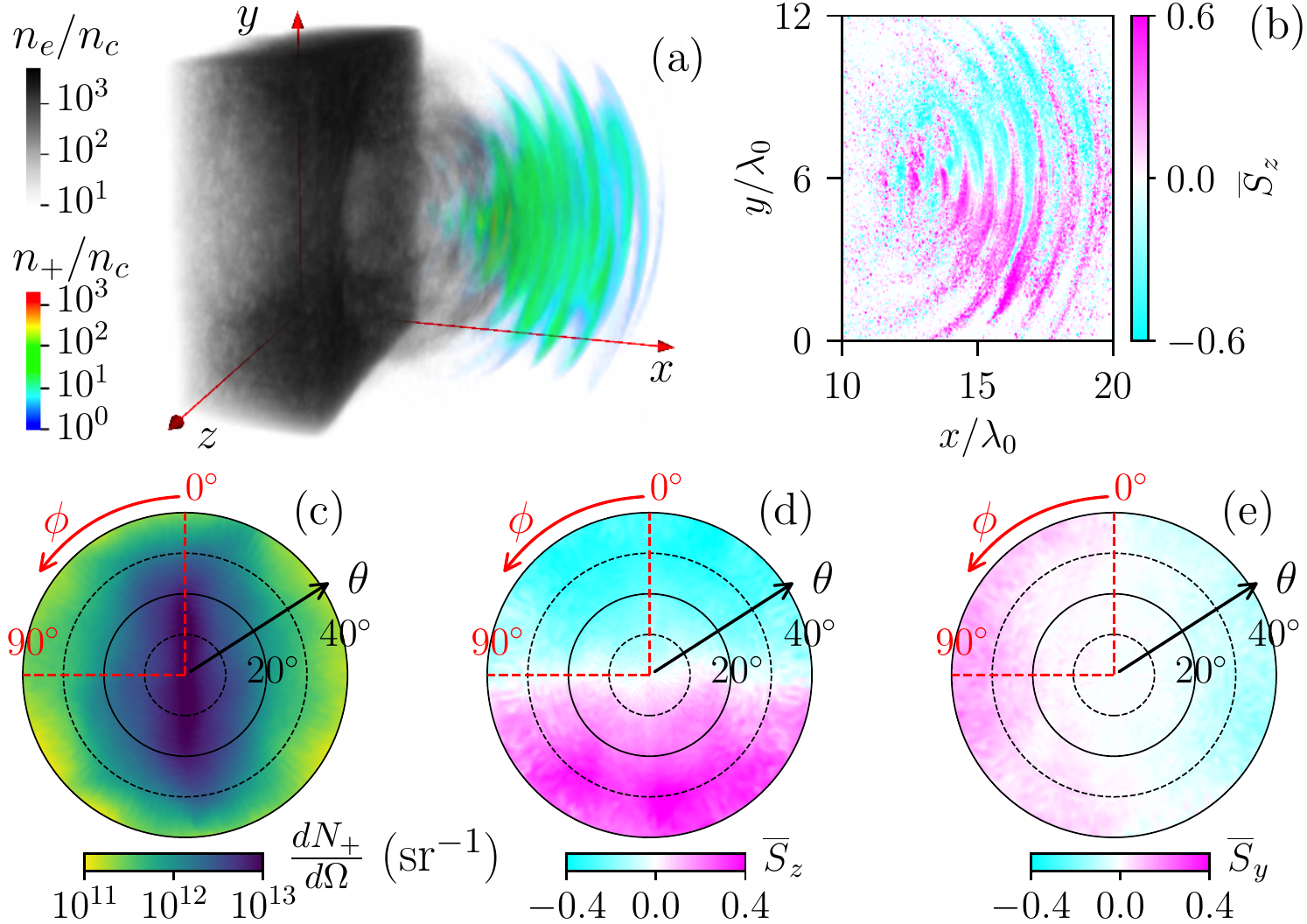}
\caption{\label{fig5}3D PIC simulation results. (a) Densities of the
electron $n_e$ and positron $n_+$. (b) A slice of positron
polarization $\overline S_z$ at the $z=6\lambda_0$ plane.  (c) Positron
number and two positron polarization components (d) $\overline
S_z$ and (e) $\overline S_y$ versus the polar angle $\theta$ and
azimuthal angle $\phi$, where the laser is polarized along $\phi=0,
180^\circ$ and propagates along $\theta=0$.}
\end{figure}

\textit{3D simulations.}---Finally, 3D PIC simulations are conducted
to validate the above 1D results. The incident laser has a
transverse profile of exp$[-(y^2+z^2)/\sigma_0^2]$ with
$\sigma_0=2.0~\mu$m (peak power of 197 PW). Other laser and target
parameters are the same as Fig.~\ref{fig1}. Figure~\ref{fig5}(a)
presents that positrons are composed of discrete bunches with a
period of half-laser-cycle, consistent with Fig.~\ref{fig2}(a).
Similar angularly-dependent polarization $\overline S_z$ can be
observed in Fig.~\ref{fig5}(b). The total positron yield is 550 nC;
among them, $>30$ nC positrons acquire a polarization above $30\%$
(angular number density $>10^{12}$ sr$^{-1}$), with the yield 5
orders of magnitude higher than laser-beam-collision schemes
\cite{Wan2020plb,Chen2019prl}. Due to multidimensional effects,
positrons are also slightly polarized in the $y$ directions
[Fig.~\ref{fig5}(e)]. This could arise from other laser field
components due to the tightly focusing. Also, we find that the
polarization is slightly weaker when the laser is obliquely
incident, and there are 22 nC positrons with the polarization above
30\% for a 30$^\circ$ incidence angle (Fig.~S8 in \cite{material}).


In conclusion, we have investigated the generation of dense
polarized positrons in a conventional setup of laser-solid
interaction in the QED regime. Over 30 nC transversely polarized
positrons with the polarization degree above 30\% can be generated
at the laser intensity $3\times10^{24}~\rm W/\rm cm^2$. The high
density and charge of such polarized positrons can meet the
requirements of future electron-positron colliders and exploration
of polarized plasma collective behavior \cite{Brodin2007njp}. The
positron polarization mechanism is robust since the laser fields
experienced by positrons are naturally asymmetric in the skin layer.
Therefore, the positron polarization could be ubiquitous in future
100-PW laser-solid experiments.

\begin{acknowledgments}
This work was supported by the National Key R\&D Program of China
(Grant No. 2018YFA0404801), National Natural Science Foundation of
China (Grant Nos. 11775302 and 11827807), the Strategic Priority
Research Program of Chinese Academy of Sciences (Grant Nos.
XDA25050300 and XDA25010300), and the Fundamental Research Funds for
the Central Universities, the Research Funds of Renmin University of
China (20XNLG01).
\end{acknowledgments}

%


\begin{thebibliography}{47}%
\makeatletter
\providecommand \@ifxundefined [1]{%
 \@ifx{#1\undefined}
}%
\providecommand \@ifnum [1]{%
 \ifnum #1\expandafter \@firstoftwo
 \else \expandafter \@secondoftwo
 \fi
}%
\providecommand \@ifx [1]{%
 \ifx #1\expandafter \@firstoftwo
 \else \expandafter \@secondoftwo
 \fi
}%
\providecommand \natexlab [1]{#1}%
\providecommand \enquote  [1]{``#1''}%
\providecommand \bibnamefont  [1]{#1}%
\providecommand \bibfnamefont [1]{#1}%
\providecommand \citenamefont [1]{#1}%
\providecommand \href@noop [0]{\@secondoftwo}%
\providecommand \href [0]{\begingroup \@sanitize@url \@href}%
\providecommand \@href[1]{\@@startlink{#1}\@@href}%
\providecommand \@@href[1]{\endgroup#1\@@endlink}%
\providecommand \@sanitize@url [0]{\catcode `\\12\catcode `\$12\catcode
  `\&12\catcode `\#12\catcode `\^12\catcode `\_12\catcode `\%12\relax}%
\providecommand \@@startlink[1]{}%
\providecommand \@@endlink[0]{}%
\providecommand \url  [0]{\begingroup\@sanitize@url \@url }%
\providecommand \@url [1]{\endgroup\@href {#1}{\urlprefix }}%
\providecommand \urlprefix  [0]{URL }%
\providecommand \Eprint [0]{\href }%
\providecommand \doibase [0]{https://doi.org/}%
\providecommand \selectlanguage [0]{\@gobble}%
\providecommand \bibinfo  [0]{\@secondoftwo}%
\providecommand \bibfield  [0]{\@secondoftwo}%
\providecommand \translation [1]{[#1]}%
\providecommand \BibitemOpen [0]{}%
\providecommand \bibitemStop [0]{}%
\providecommand \bibitemNoStop [0]{.\EOS\space}%
\providecommand \EOS [0]{\spacefactor3000\relax}%
\providecommand \BibitemShut  [1]{\csname bibitem#1\endcsname}%
\let\auto@bib@innerbib\@empty
\bibitem [{\citenamefont {Moortgat-Pick}\ \emph {et~al.}(2008)\citenamefont
  {Moortgat-Pick}, \citenamefont {Abe}, \citenamefont {Alexander},
  \citenamefont {Ananthanarayan}, \citenamefont {Babich}, \citenamefont
  {Bharadwaj}, \citenamefont {Barber}, \citenamefont {Bartl}, \citenamefont
  {Brachmann}, \citenamefont {Chen} \emph {et~al.}}]{Moortgat2008pr}%
  \BibitemOpen
  \bibfield  {author} {\bibinfo {author} {\bibfnamefont {G.}~\bibnamefont
  {Moortgat-Pick}}, \bibinfo {author} {\bibfnamefont {T.}~\bibnamefont {Abe}},
  \bibinfo {author} {\bibfnamefont {G.}~\bibnamefont {Alexander}}, \bibinfo
  {author} {\bibfnamefont {B.}~\bibnamefont {Ananthanarayan}}, \bibinfo
  {author} {\bibfnamefont {A.}~\bibnamefont {Babich}}, \bibinfo {author}
  {\bibfnamefont {V.}~\bibnamefont {Bharadwaj}}, \bibinfo {author}
  {\bibfnamefont {D.}~\bibnamefont {Barber}}, \bibinfo {author} {\bibfnamefont
  {A.}~\bibnamefont {Bartl}}, \bibinfo {author} {\bibfnamefont
  {A.}~\bibnamefont {Brachmann}}, \bibinfo {author} {\bibfnamefont
  {S.}~\bibnamefont {Chen}}, \emph {et~al.},\ }\bibfield  {title} {\bibinfo
  {title} {Polarized positrons and electrons at the linear collider},\ }\href
  {https://doi.org/https://doi.org/10.1016/j.physrep.2007.12.003} {\bibfield
  {journal} {\bibinfo  {journal} {Phys. Rep.}\ }\textbf {\bibinfo {volume}
  {460}},\ \bibinfo {pages} {131 } (\bibinfo {year} {2008})}\BibitemShut
  {NoStop}%
\bibitem [{\citenamefont {Vauth}\ and\ \citenamefont
  {List}(2016)}]{Vauth2016ijmp}%
  \BibitemOpen
  \bibfield  {author} {\bibinfo {author} {\bibfnamefont {A.}~\bibnamefont
  {Vauth}}\ and\ \bibinfo {author} {\bibfnamefont {J.}~\bibnamefont {List}},\
  }\bibfield  {title} {\bibinfo {title} {Beam polarization at the {ILC}:
  Physics case and realization},\ }\href
  {https://doi.org/10.1142/S201019451660003X} {\bibfield  {journal} {\bibinfo
  {journal} {Int. J. Mod. Phys.: Conf. Ser.}\ }\textbf {\bibinfo {volume}
  {40}},\ \bibinfo {pages} {1660003} (\bibinfo {year} {2016})}\BibitemShut
  {NoStop}%
\bibitem [{\citenamefont {Gidley}\ \emph {et~al.}(1982)\citenamefont {Gidley},
  \citenamefont {K\"oymen},\ and\ \citenamefont {Capehart}}]{Gidley1982prl}%
  \BibitemOpen
  \bibfield  {author} {\bibinfo {author} {\bibfnamefont {D.~W.}\ \bibnamefont
  {Gidley}}, \bibinfo {author} {\bibfnamefont {A.~R.}\ \bibnamefont
  {K\"oymen}},\ and\ \bibinfo {author} {\bibfnamefont {T.~W.}\ \bibnamefont
  {Capehart}},\ }\bibfield  {title} {\bibinfo {title} {Polarized low-energy
  positrons: A new probe of surface magnetism},\ }\href
  {https://doi.org/10.1103/PhysRevLett.49.1779} {\bibfield  {journal} {\bibinfo
   {journal} {Phys. Rev. Lett.}\ }\textbf {\bibinfo {volume} {49}},\ \bibinfo
  {pages} {1779} (\bibinfo {year} {1982})}\BibitemShut {NoStop}%
\bibitem [{\citenamefont {Hugenschmidt}(2016)}]{Hugenschmidt2016ssr}%
  \BibitemOpen
  \bibfield  {author} {\bibinfo {author} {\bibfnamefont {C.}~\bibnamefont
  {Hugenschmidt}},\ }\bibfield  {title} {\bibinfo {title} {Positrons in surface
  physics},\ }\href
  {https://doi.org/https://doi.org/10.1016/j.surfrep.2016.09.002} {\bibfield
  {journal} {\bibinfo  {journal} {Surf. Sci. Rep.}\ }\textbf {\bibinfo {volume}
  {71}},\ \bibinfo {pages} {547 } (\bibinfo {year} {2016})}\BibitemShut
  {NoStop}%
\bibitem [{\citenamefont {Novak}\ and\ \citenamefont
  {Kholodov}(2009)}]{Novak2009prd}%
  \BibitemOpen
  \bibfield  {author} {\bibinfo {author} {\bibfnamefont {O.~P.}\ \bibnamefont
  {Novak}}\ and\ \bibinfo {author} {\bibfnamefont {R.~I.}\ \bibnamefont
  {Kholodov}},\ }\bibfield  {title} {\bibinfo {title} {Spin-polarization
  effects in the processes of synchrotron radiation and electron-positron pair
  production by a photon in a magnetic field},\ }\href
  {https://doi.org/10.1103/PhysRevD.80.025025} {\bibfield  {journal} {\bibinfo
  {journal} {Phys. Rev. D}\ }\textbf {\bibinfo {volume} {80}},\ \bibinfo
  {pages} {025025} (\bibinfo {year} {2009})}\BibitemShut {NoStop}%
\bibitem [{\citenamefont {Sokolov}\ and\ \citenamefont
  {Ternov}(1968)}]{Sokolov1968qed}%
  \BibitemOpen
  \bibfield  {author} {\bibinfo {author} {\bibfnamefont {A.~A.}\ \bibnamefont
  {Sokolov}}\ and\ \bibinfo {author} {\bibfnamefont {I.~M.}\ \bibnamefont
  {Ternov}},\ }\href@noop {} {\emph {\bibinfo {title} {Synchrotron
  Radiation}}}\ (\bibinfo  {publisher} {Akademie-Verlag, Berlin},\ \bibinfo
  {year} {1968})\BibitemShut {NoStop}%
\bibitem [{\citenamefont {Baier}(1972)}]{Baier1972spu}%
  \BibitemOpen
  \bibfield  {author} {\bibinfo {author} {\bibfnamefont {V.~N.}\ \bibnamefont
  {Baier}},\ }\bibfield  {title} {\bibinfo {title} {Radiative polarization of
  electrons in storage rings},\ }\href
  {https://doi.org/10.1070/pu1972v014n06abeh004751} {\bibfield  {journal}
  {\bibinfo  {journal} {Sov. Phys. Usp.}\ }\textbf {\bibinfo {volume} {14}},\
  \bibinfo {pages} {695} (\bibinfo {year} {1972})}\BibitemShut {NoStop}%
\bibitem [{\citenamefont {Alexander}\ \emph {et~al.}(2008)\citenamefont
  {Alexander}, \citenamefont {Barley}, \citenamefont {Batygin}, \citenamefont
  {Berridge}, \citenamefont {Bharadwaj}, \citenamefont {Bower}, \citenamefont
  {Bugg}, \citenamefont {Decker}, \citenamefont {Dollan}, \citenamefont
  {Efremenko} \emph {et~al.}}]{Alexander2008prl}%
  \BibitemOpen
  \bibfield  {author} {\bibinfo {author} {\bibfnamefont {G.}~\bibnamefont
  {Alexander}}, \bibinfo {author} {\bibfnamefont {J.}~\bibnamefont {Barley}},
  \bibinfo {author} {\bibfnamefont {Y.}~\bibnamefont {Batygin}}, \bibinfo
  {author} {\bibfnamefont {S.}~\bibnamefont {Berridge}}, \bibinfo {author}
  {\bibfnamefont {V.}~\bibnamefont {Bharadwaj}}, \bibinfo {author}
  {\bibfnamefont {G.}~\bibnamefont {Bower}}, \bibinfo {author} {\bibfnamefont
  {W.}~\bibnamefont {Bugg}}, \bibinfo {author} {\bibfnamefont {F.-J.}\
  \bibnamefont {Decker}}, \bibinfo {author} {\bibfnamefont {R.}~\bibnamefont
  {Dollan}}, \bibinfo {author} {\bibfnamefont {Y.}~\bibnamefont {Efremenko}},
  \emph {et~al.},\ }\bibfield  {title} {\bibinfo {title} {Observation of
  polarized positrons from an undulator-based source},\ }\href
  {https://doi.org/10.1103/PhysRevLett.100.210801} {\bibfield  {journal}
  {\bibinfo  {journal} {Phys. Rev. Lett.}\ }\textbf {\bibinfo {volume} {100}},\
  \bibinfo {pages} {210801} (\bibinfo {year} {2008})}\BibitemShut {NoStop}%
\bibitem [{\citenamefont {Omori}\ \emph {et~al.}(2006)\citenamefont {Omori},
  \citenamefont {Fukuda}, \citenamefont {Hirose}, \citenamefont {Kurihara},
  \citenamefont {Kuroda}, \citenamefont {Nomura}, \citenamefont {Ohashi},
  \citenamefont {Okugi}, \citenamefont {Sakaue}, \citenamefont {Saito},
  \citenamefont {Urakawa}, \citenamefont {Washio},\ and\ \citenamefont
  {Yamazaki}}]{Omori2006prl}%
  \BibitemOpen
  \bibfield  {author} {\bibinfo {author} {\bibfnamefont {T.}~\bibnamefont
  {Omori}}, \bibinfo {author} {\bibfnamefont {M.}~\bibnamefont {Fukuda}},
  \bibinfo {author} {\bibfnamefont {T.}~\bibnamefont {Hirose}}, \bibinfo
  {author} {\bibfnamefont {Y.}~\bibnamefont {Kurihara}}, \bibinfo {author}
  {\bibfnamefont {R.}~\bibnamefont {Kuroda}}, \bibinfo {author} {\bibfnamefont
  {M.}~\bibnamefont {Nomura}}, \bibinfo {author} {\bibfnamefont
  {A.}~\bibnamefont {Ohashi}}, \bibinfo {author} {\bibfnamefont
  {T.}~\bibnamefont {Okugi}}, \bibinfo {author} {\bibfnamefont
  {K.}~\bibnamefont {Sakaue}}, \bibinfo {author} {\bibfnamefont
  {T.}~\bibnamefont {Saito}}, \bibinfo {author} {\bibfnamefont
  {J.}~\bibnamefont {Urakawa}}, \bibinfo {author} {\bibfnamefont
  {M.}~\bibnamefont {Washio}},\ and\ \bibinfo {author} {\bibfnamefont
  {I.}~\bibnamefont {Yamazaki}},\ }\bibfield  {title} {\bibinfo {title}
  {Efficient propagation of polarization from laser photons to positrons
  through {C}ompton scattering and electron-positron pair creation},\ }\href
  {https://doi.org/10.1103/PhysRevLett.96.114801} {\bibfield  {journal}
  {\bibinfo  {journal} {Phys. Rev. Lett.}\ }\textbf {\bibinfo {volume} {96}},\
  \bibinfo {pages} {114801} (\bibinfo {year} {2006})}\BibitemShut {NoStop}%
\bibitem [{\citenamefont {Abbott}\ \emph {et~al.}(2016)\citenamefont {Abbott},
  \citenamefont {Adderley}, \citenamefont {Adeyemi}, \citenamefont {Aguilera},
  \citenamefont {Ali}, \citenamefont {Areti}, \citenamefont {Baylac},
  \citenamefont {Benesch}, \citenamefont {Bosson}, \citenamefont {Cade} \emph
  {et~al.}}]{Abbott2016prl}%
  \BibitemOpen
  \bibfield  {author} {\bibinfo {author} {\bibfnamefont {D.}~\bibnamefont
  {Abbott}}, \bibinfo {author} {\bibfnamefont {P.}~\bibnamefont {Adderley}},
  \bibinfo {author} {\bibfnamefont {A.}~\bibnamefont {Adeyemi}}, \bibinfo
  {author} {\bibfnamefont {P.}~\bibnamefont {Aguilera}}, \bibinfo {author}
  {\bibfnamefont {M.}~\bibnamefont {Ali}}, \bibinfo {author} {\bibfnamefont
  {H.}~\bibnamefont {Areti}}, \bibinfo {author} {\bibfnamefont
  {M.}~\bibnamefont {Baylac}}, \bibinfo {author} {\bibfnamefont
  {J.}~\bibnamefont {Benesch}}, \bibinfo {author} {\bibfnamefont
  {G.}~\bibnamefont {Bosson}}, \bibinfo {author} {\bibfnamefont
  {B.}~\bibnamefont {Cade}}, \emph {et~al.} (\bibinfo {collaboration} {PEPPo
  Collaboration}),\ }\bibfield  {title} {\bibinfo {title} {Production of highly
  polarized positrons using polarized electrons at {M}e{V} energies},\ }\href
  {https://doi.org/10.1103/PhysRevLett.116.214801} {\bibfield  {journal}
  {\bibinfo  {journal} {Phys. Rev. Lett.}\ }\textbf {\bibinfo {volume} {116}},\
  \bibinfo {pages} {214801} (\bibinfo {year} {2016})}\BibitemShut {NoStop}%
\bibitem [{\citenamefont {Berestetskii}\ \emph {et~al.}(1982)\citenamefont
  {Berestetskii}, \citenamefont {Lifshitz},\ and\ \citenamefont
  {Pitaevskii}}]{Berestetskii1982qed}%
  \BibitemOpen
  \bibfield  {author} {\bibinfo {author} {\bibfnamefont {V.~B.}\ \bibnamefont
  {Berestetskii}}, \bibinfo {author} {\bibfnamefont {E.~M.}\ \bibnamefont
  {Lifshitz}},\ and\ \bibinfo {author} {\bibfnamefont {L.~P.}\ \bibnamefont
  {Pitaevskii}},\ }\href@noop {} {\emph {\bibinfo {title} {Quantum
  Electrodynamics}}}\ (\bibinfo  {publisher} {Elsevier Butterworth-Heinemann,
  Oxford},\ \bibinfo {year} {1982})\BibitemShut {NoStop}%
\bibitem [{\citenamefont {Baier}\ \emph {et~al.}(1998)\citenamefont {Baier},
  \citenamefont {Katkov},\ and\ \citenamefont {Strakhovenko}}]{Baier1998qed}%
  \BibitemOpen
  \bibfield  {author} {\bibinfo {author} {\bibfnamefont {V.~N.}\ \bibnamefont
  {Baier}}, \bibinfo {author} {\bibfnamefont {V.}~\bibnamefont {Katkov}},\ and\
  \bibinfo {author} {\bibfnamefont {V.~M.}\ \bibnamefont {Strakhovenko}},\
  }\href@noop {} {\emph {\bibinfo {title} {Electromagnetic Processes at High
  Energies in Oriented Single Crystals}}}\ (\bibinfo  {publisher} {World
  Scientific, Singapore},\ \bibinfo {year} {1998})\BibitemShut {NoStop}%
\bibitem [{\citenamefont {Ritus}(1985)}]{Ritus1985jslr}%
  \BibitemOpen
  \bibfield  {author} {\bibinfo {author} {\bibfnamefont {V.~I.}\ \bibnamefont
  {Ritus}},\ }\bibfield  {title} {\bibinfo {title} {Quantum effects of the
  interaction of elementary particles with an intense electromagnetic field},\
  }\href {https://doi.org/https://doi.org/10.1007/BF01120220} {\bibfield
  {journal} {\bibinfo  {journal} {J. Sov. Laser Res.}\ }\textbf {\bibinfo
  {volume} {6}},\ \bibinfo {pages} {497} (\bibinfo {year} {1985})}\BibitemShut
  {NoStop}%
\bibitem [{\citenamefont {Di~Piazza}\ \emph {et~al.}(2012)\citenamefont
  {Di~Piazza}, \citenamefont {M\"uller}, \citenamefont {Hatsagortsyan},\ and\
  \citenamefont {Keitel}}]{piazza2012rmp}%
  \BibitemOpen
  \bibfield  {author} {\bibinfo {author} {\bibfnamefont {A.}~\bibnamefont
  {Di~Piazza}}, \bibinfo {author} {\bibfnamefont {C.}~\bibnamefont {M\"uller}},
  \bibinfo {author} {\bibfnamefont {K.~Z.}\ \bibnamefont {Hatsagortsyan}},\
  and\ \bibinfo {author} {\bibfnamefont {C.~H.}\ \bibnamefont {Keitel}},\
  }\bibfield  {title} {\bibinfo {title} {Extremely high-intensity laser
  interactions with fundamental quantum systems},\ }\href
  {https://doi.org/10.1103/RevModPhys.84.1177} {\bibfield  {journal} {\bibinfo
  {journal} {Rev. Mod. Phys.}\ }\textbf {\bibinfo {volume} {84}},\ \bibinfo
  {pages} {1177} (\bibinfo {year} {2012})}\BibitemShut {NoStop}%
\bibitem [{\citenamefont {Gonoskov}\ \emph {et~al.}()\citenamefont {Gonoskov},
  \citenamefont {Blackburn}, \citenamefont {Marklund},\ and\ \citenamefont
  {Bulanov}}]{Gonoskov2021arxiv}%
  \BibitemOpen
  \bibfield  {author} {\bibinfo {author} {\bibfnamefont {A.}~\bibnamefont
  {Gonoskov}}, \bibinfo {author} {\bibfnamefont {T.~G.}\ \bibnamefont
  {Blackburn}}, \bibinfo {author} {\bibfnamefont {M.}~\bibnamefont
  {Marklund}},\ and\ \bibinfo {author} {\bibfnamefont {S.~S.}\ \bibnamefont
  {Bulanov}},\ }\href@noop {} {\bibinfo {title} {Charged particle motion and
  radiation in strong electromagnetic fields}},\ \Eprint
  {https://arxiv.org/abs/2107.02161} {arXiv:2107.02161} \BibitemShut {NoStop}%
\bibitem [{\citenamefont {Danson}\ \emph {et~al.}(2019)\citenamefont {Danson},
  \citenamefont {Haefner}, \citenamefont {Bromage}, \citenamefont {Butcher},
  \citenamefont {Chanteloup}, \citenamefont {Chowdhury}, \citenamefont
  {Galvanauskas}, \citenamefont {Gizzi}, \citenamefont {Hein}, \citenamefont
  {Hillier},\ and\ \citenamefont {et~al.}}]{Danson2019hplse}%
  \BibitemOpen
  \bibfield  {author} {\bibinfo {author} {\bibfnamefont {C.~N.}\ \bibnamefont
  {Danson}}, \bibinfo {author} {\bibfnamefont {C.}~\bibnamefont {Haefner}},
  \bibinfo {author} {\bibfnamefont {J.}~\bibnamefont {Bromage}}, \bibinfo
  {author} {\bibfnamefont {T.}~\bibnamefont {Butcher}}, \bibinfo {author}
  {\bibfnamefont {J.-C.~F.}\ \bibnamefont {Chanteloup}}, \bibinfo {author}
  {\bibfnamefont {E.~A.}\ \bibnamefont {Chowdhury}}, \bibinfo {author}
  {\bibfnamefont {A.}~\bibnamefont {Galvanauskas}}, \bibinfo {author}
  {\bibfnamefont {L.~A.}\ \bibnamefont {Gizzi}}, \bibinfo {author}
  {\bibfnamefont {J.}~\bibnamefont {Hein}}, \bibinfo {author} {\bibfnamefont
  {D.~I.}\ \bibnamefont {Hillier}},\ and\ \bibinfo {author} {\bibnamefont
  {et~al.}},\ }\bibfield  {title} {\bibinfo {title} {Petawatt and exawatt class
  lasers worldwide},\ }\href {https://doi.org/10.1017/hpl.2019.36} {\bibfield
  {journal} {\bibinfo  {journal} {High Power Laser Sci. Eng.}\ }\textbf
  {\bibinfo {volume} {7}},\ \bibinfo {pages} {e54} (\bibinfo {year}
  {2019})}\BibitemShut {NoStop}%
\bibitem [{\citenamefont {Yoon}\ \emph {et~al.}(2021)\citenamefont {Yoon},
  \citenamefont {Kim}, \citenamefont {Choi}, \citenamefont {Sung},
  \citenamefont {Lee}, \citenamefont {Lee},\ and\ \citenamefont
  {Nam}}]{Yoon2021optica}%
  \BibitemOpen
  \bibfield  {author} {\bibinfo {author} {\bibfnamefont {J.~W.}\ \bibnamefont
  {Yoon}}, \bibinfo {author} {\bibfnamefont {Y.~G.}\ \bibnamefont {Kim}},
  \bibinfo {author} {\bibfnamefont {I.~W.}\ \bibnamefont {Choi}}, \bibinfo
  {author} {\bibfnamefont {J.~H.}\ \bibnamefont {Sung}}, \bibinfo {author}
  {\bibfnamefont {H.~W.}\ \bibnamefont {Lee}}, \bibinfo {author} {\bibfnamefont
  {S.~K.}\ \bibnamefont {Lee}},\ and\ \bibinfo {author} {\bibfnamefont {C.~H.}\
  \bibnamefont {Nam}},\ }\href {https://doi.org/10.1364/OPTICA.420520}
  {\bibfield  {journal} {\bibinfo  {journal} {Optica}\ }\textbf {\bibinfo
  {volume} {8}},\ \bibinfo {pages} {630} (\bibinfo {year} {2021})}\BibitemShut
  {NoStop}%
\bibitem [{ELI()}]{ELI}%
  \BibitemOpen
  \href@noop {} {}\bibinfo {note} {Extreme Light Infrastructure (ELI),
  \url{https://eli-laser.eu}}\BibitemShut {NoStop}%
\bibitem [{\citenamefont {Cartlidge}(2018)}]{Cartlidge2018science}%
  \BibitemOpen
  \bibfield  {author} {\bibinfo {author} {\bibfnamefont {E.}~\bibnamefont
  {Cartlidge}},\ }\bibfield  {title} {\bibinfo {title} {The light fantastic},\
  }\href {https://doi.org/10.1126/science.359.6374.382} {\bibfield  {journal}
  {\bibinfo  {journal} {Science}\ }\textbf {\bibinfo {volume} {359}},\ \bibinfo
  {pages} {382} (\bibinfo {year} {2018})}\BibitemShut {NoStop}%
\bibitem [{\citenamefont {Shao}\ \emph {et~al.}(2020)\citenamefont {Shao},
  \citenamefont {Li}, \citenamefont {Peng}, \citenamefont {Wang}, \citenamefont
  {Qian}, \citenamefont {Leng},\ and\ \citenamefont {Li}}]{Shao2020ol}%
  \BibitemOpen
  \bibfield  {author} {\bibinfo {author} {\bibfnamefont {B.}~\bibnamefont
  {Shao}}, \bibinfo {author} {\bibfnamefont {Y.}~\bibnamefont {Li}}, \bibinfo
  {author} {\bibfnamefont {Y.}~\bibnamefont {Peng}}, \bibinfo {author}
  {\bibfnamefont {P.}~\bibnamefont {Wang}}, \bibinfo {author} {\bibfnamefont
  {J.}~\bibnamefont {Qian}}, \bibinfo {author} {\bibfnamefont {Y.}~\bibnamefont
  {Leng}},\ and\ \bibinfo {author} {\bibfnamefont {R.}~\bibnamefont {Li}},\
  }\bibfield  {title} {\bibinfo {title} {Broad-bandwidth high-temporal-contrast
  carrier-envelope-phase-stabilized laser seed for 100 {PW} lasers},\ }\href
  {https://doi.org/10.1364/OL.390110} {\bibfield  {journal} {\bibinfo
  {journal} {Opt. Lett.}\ }\textbf {\bibinfo {volume} {45}},\ \bibinfo {pages}
  {2215} (\bibinfo {year} {2020})}\BibitemShut {NoStop}%
\bibitem [{XCE()}]{XCELS}%
  \BibitemOpen
  \href@noop {} {}\bibinfo {note} {Exawatt Center for Extreme Light Studies
  (XCELS), \url{https://xcels.iapras.ru}}\BibitemShut {NoStop}%
\bibitem [{\citenamefont {{Cartlidge}}(2017)}]{Cartlidge2017science}%
  \BibitemOpen
  \bibfield  {author} {\bibinfo {author} {\bibfnamefont {E.}~\bibnamefont
  {{Cartlidge}}},\ }\bibfield  {title} {\bibinfo {title} {{Eastern Europe's
  laser centers will debut without a star}},\ }\href
  {https://doi.org/10.1126/science.355.6327.785} {\bibfield  {journal}
  {\bibinfo  {journal} {Science}\ }\textbf {\bibinfo {volume} {355}},\ \bibinfo
  {pages} {785} (\bibinfo {year} {2017})}\BibitemShut {NoStop}%
\bibitem [{\citenamefont {Reiss}(1962)}]{Reiss1962jmp}%
  \BibitemOpen
  \bibfield  {author} {\bibinfo {author} {\bibfnamefont {H.~R.}\ \bibnamefont
  {Reiss}},\ }\bibfield  {title} {\bibinfo {title} {Absorption of light by
  light},\ }\href {https://doi.org/10.1063/1.1703787} {\bibfield  {journal}
  {\bibinfo  {journal} {J. Math. Phys.}\ }\textbf {\bibinfo {volume} {3}},\
  \bibinfo {pages} {59} (\bibinfo {year} {1962})}\BibitemShut {NoStop}%
\bibitem [{\citenamefont {Sokolov}\ \emph {et~al.}(2010)\citenamefont
  {Sokolov}, \citenamefont {Naumova}, \citenamefont {Nees},\ and\ \citenamefont
  {Mourou}}]{Sokolov2010prl}%
  \BibitemOpen
  \bibfield  {author} {\bibinfo {author} {\bibfnamefont {I.~V.}\ \bibnamefont
  {Sokolov}}, \bibinfo {author} {\bibfnamefont {N.~M.}\ \bibnamefont
  {Naumova}}, \bibinfo {author} {\bibfnamefont {J.~A.}\ \bibnamefont {Nees}},\
  and\ \bibinfo {author} {\bibfnamefont {G.~A.}\ \bibnamefont {Mourou}},\
  }\bibfield  {title} {\bibinfo {title} {Pair creation in {QED}-strong pulsed
  laser fields interacting with electron beams},\ }\href
  {https://doi.org/10.1103/PhysRevLett.105.195005} {\bibfield  {journal}
  {\bibinfo  {journal} {Phys. Rev. Lett.}\ }\textbf {\bibinfo {volume} {105}},\
  \bibinfo {pages} {195005} (\bibinfo {year} {2010})}\BibitemShut {NoStop}%
\bibitem [{\citenamefont {Blackburn}\ \emph {et~al.}(2017)\citenamefont
  {Blackburn}, \citenamefont {Ilderton}, \citenamefont {Murphy},\ and\
  \citenamefont {Marklund}}]{Blackburn2017pra}%
  \BibitemOpen
  \bibfield  {author} {\bibinfo {author} {\bibfnamefont {T.~G.}\ \bibnamefont
  {Blackburn}}, \bibinfo {author} {\bibfnamefont {A.}~\bibnamefont {Ilderton}},
  \bibinfo {author} {\bibfnamefont {C.~D.}\ \bibnamefont {Murphy}},\ and\
  \bibinfo {author} {\bibfnamefont {M.}~\bibnamefont {Marklund}},\ }\bibfield
  {title} {\bibinfo {title} {Scaling laws for positron production in
  laser--electron-beam collisions},\ }\href
  {https://doi.org/10.1103/PhysRevA.96.022128} {\bibfield  {journal} {\bibinfo
  {journal} {Phys. Rev. A}\ }\textbf {\bibinfo {volume} {96}},\ \bibinfo
  {pages} {022128} (\bibinfo {year} {2017})}\BibitemShut {NoStop}%
\bibitem [{\citenamefont {Wan}\ \emph {et~al.}(2020)\citenamefont {Wan},
  \citenamefont {Shaisultanov}, \citenamefont {Li}, \citenamefont
  {Hatsagortsyan}, \citenamefont {Keitel},\ and\ \citenamefont
  {Li}}]{Wan2020plb}%
  \BibitemOpen
  \bibfield  {author} {\bibinfo {author} {\bibfnamefont {F.}~\bibnamefont
  {Wan}}, \bibinfo {author} {\bibfnamefont {R.}~\bibnamefont {Shaisultanov}},
  \bibinfo {author} {\bibfnamefont {Y.-F.}\ \bibnamefont {Li}}, \bibinfo
  {author} {\bibfnamefont {K.~Z.}\ \bibnamefont {Hatsagortsyan}}, \bibinfo
  {author} {\bibfnamefont {C.~H.}\ \bibnamefont {Keitel}},\ and\ \bibinfo
  {author} {\bibfnamefont {J.-X.}\ \bibnamefont {Li}},\ }\bibfield  {title}
  {\bibinfo {title} {Ultrarelativistic polarized positron jets via collision of
  electron and ultraintense laser beams},\ }\href
  {https://doi.org/https://doi.org/10.1016/j.physletb.2019.135120} {\bibfield
  {journal} {\bibinfo  {journal} {Phys. Lett. B}\ }\textbf {\bibinfo {volume}
  {800}},\ \bibinfo {pages} {135120} (\bibinfo {year} {2020})}\BibitemShut
  {NoStop}%
\bibitem [{\citenamefont {Chen}\ \emph {et~al.}(2019)\citenamefont {Chen},
  \citenamefont {He}, \citenamefont {Shaisultanov}, \citenamefont
  {Hatsagortsyan},\ and\ \citenamefont {Keitel}}]{Chen2019prl}%
  \BibitemOpen
  \bibfield  {author} {\bibinfo {author} {\bibfnamefont {Y.-Y.}\ \bibnamefont
  {Chen}}, \bibinfo {author} {\bibfnamefont {P.-L.}\ \bibnamefont {He}},
  \bibinfo {author} {\bibfnamefont {R.}~\bibnamefont {Shaisultanov}}, \bibinfo
  {author} {\bibfnamefont {K.~Z.}\ \bibnamefont {Hatsagortsyan}},\ and\
  \bibinfo {author} {\bibfnamefont {C.~H.}\ \bibnamefont {Keitel}},\ }\bibfield
   {title} {\bibinfo {title} {Polarized positron beams via intense two-color
  laser pulses},\ }\href {https://doi.org/10.1103/PhysRevLett.123.174801}
  {\bibfield  {journal} {\bibinfo  {journal} {Phys. Rev. Lett.}\ }\textbf
  {\bibinfo {volume} {123}},\ \bibinfo {pages} {174801} (\bibinfo {year}
  {2019})}\BibitemShut {NoStop}%
\bibitem [{\citenamefont {Tien}\ \emph {et~al.}(1999)\citenamefont {Tien},
  \citenamefont {Backus}, \citenamefont {Kapteyn}, \citenamefont {Murnane},\
  and\ \citenamefont {Mourou}}]{Tien1999prl}%
  \BibitemOpen
  \bibfield  {author} {\bibinfo {author} {\bibfnamefont {A.-C.}\ \bibnamefont
  {Tien}}, \bibinfo {author} {\bibfnamefont {S.}~\bibnamefont {Backus}},
  \bibinfo {author} {\bibfnamefont {H.}~\bibnamefont {Kapteyn}}, \bibinfo
  {author} {\bibfnamefont {M.}~\bibnamefont {Murnane}},\ and\ \bibinfo {author}
  {\bibfnamefont {G.}~\bibnamefont {Mourou}},\ }\bibfield  {title} {\bibinfo
  {title} {Short-pulse laser damage in transparent materials as a function of
  pulse duration},\ }\href {https://doi.org/10.1103/PhysRevLett.82.3883}
  {\bibfield  {journal} {\bibinfo  {journal} {Phys. Rev. Lett.}\ }\textbf
  {\bibinfo {volume} {82}},\ \bibinfo {pages} {3883} (\bibinfo {year}
  {1999})}\BibitemShut {NoStop}%
\bibitem [{\citenamefont {Nerush}\ \emph {et~al.}(2011)\citenamefont {Nerush},
  \citenamefont {Kostyukov}, \citenamefont {Fedotov}, \citenamefont {Narozhny},
  \citenamefont {Elkina},\ and\ \citenamefont {Ruhl}}]{Nerush2011prl}%
  \BibitemOpen
  \bibfield  {author} {\bibinfo {author} {\bibfnamefont {E.~N.}\ \bibnamefont
  {Nerush}}, \bibinfo {author} {\bibfnamefont {I.~Y.}\ \bibnamefont
  {Kostyukov}}, \bibinfo {author} {\bibfnamefont {A.~M.}\ \bibnamefont
  {Fedotov}}, \bibinfo {author} {\bibfnamefont {N.~B.}\ \bibnamefont
  {Narozhny}}, \bibinfo {author} {\bibfnamefont {N.~V.}\ \bibnamefont
  {Elkina}},\ and\ \bibinfo {author} {\bibfnamefont {H.}~\bibnamefont {Ruhl}},\
  }\bibfield  {title} {\bibinfo {title} {Laser field absorption in
  self-generated electron-positron pair plasma},\ }\href
  {https://doi.org/10.1103/PhysRevLett.106.035001} {\bibfield  {journal}
  {\bibinfo  {journal} {Phys. Rev. Lett.}\ }\textbf {\bibinfo {volume} {106}},\
  \bibinfo {pages} {035001} (\bibinfo {year} {2011})}\BibitemShut {NoStop}%
\bibitem [{\citenamefont {Zhu}\ \emph {et~al.}(2016)\citenamefont {Zhu},
  \citenamefont {Yu}, \citenamefont {Sheng}, \citenamefont {Yin}, \citenamefont
  {Turcu},\ and\ \citenamefont {Pukhov}}]{Zhu2016nc}%
  \BibitemOpen
  \bibfield  {author} {\bibinfo {author} {\bibfnamefont {X.-L.}\ \bibnamefont
  {Zhu}}, \bibinfo {author} {\bibfnamefont {T.-P.}\ \bibnamefont {Yu}},
  \bibinfo {author} {\bibfnamefont {Z.-M.}\ \bibnamefont {Sheng}}, \bibinfo
  {author} {\bibfnamefont {Y.}~\bibnamefont {Yin}}, \bibinfo {author}
  {\bibfnamefont {I.~C.~E.}\ \bibnamefont {Turcu}},\ and\ \bibinfo {author}
  {\bibfnamefont {A.}~\bibnamefont {Pukhov}},\ }\bibfield  {title} {\bibinfo
  {title} {Dense {G}ev electron-positron pairs generated by lasers in
  near-critical-density plasmas},\ }\href
  {https://doi.org/https://doi.org/10.1038/ncomms13686} {\bibfield  {journal}
  {\bibinfo  {journal} {Nat. Commun.}\ }\textbf {\bibinfo {volume} {7}},\
  \bibinfo {pages} {13686} (\bibinfo {year} {2016})}\BibitemShut {NoStop}%
\bibitem [{\citenamefont {Grismayer}\ \emph {et~al.}(2017)\citenamefont
  {Grismayer}, \citenamefont {Vranic}, \citenamefont {Martins}, \citenamefont
  {Fonseca},\ and\ \citenamefont {Silva}}]{Grismayer2017pre}%
  \BibitemOpen
  \bibfield  {author} {\bibinfo {author} {\bibfnamefont {T.}~\bibnamefont
  {Grismayer}}, \bibinfo {author} {\bibfnamefont {M.}~\bibnamefont {Vranic}},
  \bibinfo {author} {\bibfnamefont {J.~L.}\ \bibnamefont {Martins}}, \bibinfo
  {author} {\bibfnamefont {R.~A.}\ \bibnamefont {Fonseca}},\ and\ \bibinfo
  {author} {\bibfnamefont {L.~O.}\ \bibnamefont {Silva}},\ }\bibfield  {title}
  {\bibinfo {title} {Seeded {QED} cascades in counterpropagating laser
  pulses},\ }\href {https://doi.org/10.1103/PhysRevE.95.023210} {\bibfield
  {journal} {\bibinfo  {journal} {Phys. Rev. E}\ }\textbf {\bibinfo {volume}
  {95}},\ \bibinfo {pages} {023210} (\bibinfo {year} {2017})}\BibitemShut
  {NoStop}%
\bibitem [{\citenamefont {Ridgers}\ \emph {et~al.}(2012)\citenamefont
  {Ridgers}, \citenamefont {Brady}, \citenamefont {Duclous}, \citenamefont
  {Kirk}, \citenamefont {Bennett}, \citenamefont {Arber}, \citenamefont
  {Robinson},\ and\ \citenamefont {Bell}}]{Ridgers2012prl}%
  \BibitemOpen
  \bibfield  {author} {\bibinfo {author} {\bibfnamefont {C.~P.}\ \bibnamefont
  {Ridgers}}, \bibinfo {author} {\bibfnamefont {C.~S.}\ \bibnamefont {Brady}},
  \bibinfo {author} {\bibfnamefont {R.}~\bibnamefont {Duclous}}, \bibinfo
  {author} {\bibfnamefont {J.~G.}\ \bibnamefont {Kirk}}, \bibinfo {author}
  {\bibfnamefont {K.}~\bibnamefont {Bennett}}, \bibinfo {author} {\bibfnamefont
  {T.~D.}\ \bibnamefont {Arber}}, \bibinfo {author} {\bibfnamefont {A.~P.~L.}\
  \bibnamefont {Robinson}},\ and\ \bibinfo {author} {\bibfnamefont {A.~R.}\
  \bibnamefont {Bell}},\ }\bibfield  {title} {\bibinfo {title} {Dense
  electron-positron plasmas and ultraintense $\ensuremath{\gamma}$ rays from
  laser-irradiated solids},\ }\href
  {https://doi.org/10.1103/PhysRevLett.108.165006} {\bibfield  {journal}
  {\bibinfo  {journal} {Phys. Rev. Lett.}\ }\textbf {\bibinfo {volume} {108}},\
  \bibinfo {pages} {165006} (\bibinfo {year} {2012})}\BibitemShut {NoStop}%
\bibitem [{\citenamefont {Kostyukov}\ and\ \citenamefont
  {Nerush}(2016)}]{Kostyukov2016pop}%
  \BibitemOpen
  \bibfield  {author} {\bibinfo {author} {\bibfnamefont {I.~Y.}\ \bibnamefont
  {Kostyukov}}\ and\ \bibinfo {author} {\bibfnamefont {E.~N.}\ \bibnamefont
  {Nerush}},\ }\bibfield  {title} {\bibinfo {title} {Production and dynamics of
  positrons in ultrahigh intensity laser-foil interactions},\ }\href
  {https://doi.org/10.1063/1.4962567} {\bibfield  {journal} {\bibinfo
  {journal} {Phys. Plasmas}\ }\textbf {\bibinfo {volume} {23}},\ \bibinfo
  {pages} {093119} (\bibinfo {year} {2016})}\BibitemShut {NoStop}%
\bibitem [{\citenamefont {Wang}\ \emph {et~al.}(2017)\citenamefont {Wang},
  \citenamefont {Gibbon}, \citenamefont {Sheng}, \citenamefont {Li},\ and\
  \citenamefont {Zhang}}]{Wang2017pre}%
  \BibitemOpen
  \bibfield  {author} {\bibinfo {author} {\bibfnamefont {W.-M.}\ \bibnamefont
  {Wang}}, \bibinfo {author} {\bibfnamefont {P.}~\bibnamefont {Gibbon}},
  \bibinfo {author} {\bibfnamefont {Z.-M.}\ \bibnamefont {Sheng}}, \bibinfo
  {author} {\bibfnamefont {Y.-T.}\ \bibnamefont {Li}},\ and\ \bibinfo {author}
  {\bibfnamefont {J.}~\bibnamefont {Zhang}},\ }\bibfield  {title} {\bibinfo
  {title} {Laser opacity in underdense preplasma of solid targets due to
  quantum electrodynamics effects},\ }\href
  {https://doi.org/10.1103/PhysRevE.96.013201} {\bibfield  {journal} {\bibinfo
  {journal} {Phys. Rev. E}\ }\textbf {\bibinfo {volume} {96}},\ \bibinfo
  {pages} {013201} (\bibinfo {year} {2017})}\BibitemShut {NoStop}%
\bibitem [{\citenamefont {Elkina}\ \emph {et~al.}(2011)\citenamefont {Elkina},
  \citenamefont {Fedotov}, \citenamefont {Kostyukov}, \citenamefont {Legkov},
  \citenamefont {Narozhny}, \citenamefont {Nerush},\ and\ \citenamefont
  {Ruhl}}]{Elkina2011prab}%
  \BibitemOpen
  \bibfield  {author} {\bibinfo {author} {\bibfnamefont {N.~V.}\ \bibnamefont
  {Elkina}}, \bibinfo {author} {\bibfnamefont {A.~M.}\ \bibnamefont {Fedotov}},
  \bibinfo {author} {\bibfnamefont {I.~Y.}\ \bibnamefont {Kostyukov}}, \bibinfo
  {author} {\bibfnamefont {M.~V.}\ \bibnamefont {Legkov}}, \bibinfo {author}
  {\bibfnamefont {N.~B.}\ \bibnamefont {Narozhny}}, \bibinfo {author}
  {\bibfnamefont {E.~N.}\ \bibnamefont {Nerush}},\ and\ \bibinfo {author}
  {\bibfnamefont {H.}~\bibnamefont {Ruhl}},\ }\bibfield  {title} {\bibinfo
  {title} {{QED} cascades induced by circularly polarized laser fields},\
  }\href {https://doi.org/10.1103/PhysRevSTAB.14.054401} {\bibfield  {journal}
  {\bibinfo  {journal} {Phys. Rev. ST Accel. Beams}\ }\textbf {\bibinfo
  {volume} {14}},\ \bibinfo {pages} {054401} (\bibinfo {year}
  {2011})}\BibitemShut {NoStop}%
\bibitem [{\citenamefont {Ridgers}\ \emph {et~al.}(2014)\citenamefont
  {Ridgers}, \citenamefont {Kirk}, \citenamefont {Duclous}, \citenamefont
  {Blackburn}, \citenamefont {Brady}, \citenamefont {Bennett}, \citenamefont
  {Arber},\ and\ \citenamefont {Bell}}]{Ridgers2014jcp}%
  \BibitemOpen
  \bibfield  {author} {\bibinfo {author} {\bibfnamefont {C.}~\bibnamefont
  {Ridgers}}, \bibinfo {author} {\bibfnamefont {J.}~\bibnamefont {Kirk}},
  \bibinfo {author} {\bibfnamefont {R.}~\bibnamefont {Duclous}}, \bibinfo
  {author} {\bibfnamefont {T.}~\bibnamefont {Blackburn}}, \bibinfo {author}
  {\bibfnamefont {C.}~\bibnamefont {Brady}}, \bibinfo {author} {\bibfnamefont
  {K.}~\bibnamefont {Bennett}}, \bibinfo {author} {\bibfnamefont
  {T.}~\bibnamefont {Arber}},\ and\ \bibinfo {author} {\bibfnamefont
  {A.}~\bibnamefont {Bell}},\ }\bibfield  {title} {\bibinfo {title} {Modelling
  gamma-ray photon emission and pair production in high-intensity
  laser–matter interactions},\ }\href
  {https://doi.org/https://doi.org/10.1016/j.jcp.2013.12.007} {\bibfield
  {journal} {\bibinfo  {journal} {J. Comput. Phys.}\ }\textbf {\bibinfo
  {volume} {260}},\ \bibinfo {pages} {273 } (\bibinfo {year}
  {2014})}\BibitemShut {NoStop}%
\bibitem [{\citenamefont {Gonoskov}\ \emph {et~al.}(2015)\citenamefont
  {Gonoskov}, \citenamefont {Bastrakov}, \citenamefont {Efimenko},
  \citenamefont {Ilderton}, \citenamefont {Marklund}, \citenamefont {Meyerov},
  \citenamefont {Muraviev}, \citenamefont {Sergeev}, \citenamefont {Surmin},\
  and\ \citenamefont {Wallin}}]{Gonoskov2015pre}%
  \BibitemOpen
  \bibfield  {author} {\bibinfo {author} {\bibfnamefont {A.}~\bibnamefont
  {Gonoskov}}, \bibinfo {author} {\bibfnamefont {S.}~\bibnamefont {Bastrakov}},
  \bibinfo {author} {\bibfnamefont {E.}~\bibnamefont {Efimenko}}, \bibinfo
  {author} {\bibfnamefont {A.}~\bibnamefont {Ilderton}}, \bibinfo {author}
  {\bibfnamefont {M.}~\bibnamefont {Marklund}}, \bibinfo {author}
  {\bibfnamefont {I.}~\bibnamefont {Meyerov}}, \bibinfo {author} {\bibfnamefont
  {A.}~\bibnamefont {Muraviev}}, \bibinfo {author} {\bibfnamefont
  {A.}~\bibnamefont {Sergeev}}, \bibinfo {author} {\bibfnamefont
  {I.}~\bibnamefont {Surmin}},\ and\ \bibinfo {author} {\bibfnamefont
  {E.}~\bibnamefont {Wallin}},\ }\bibfield  {title} {\bibinfo {title} {Extended
  particle-in-cell schemes for physics in ultrastrong laser fields: Review and
  developments},\ }\href {https://doi.org/10.1103/PhysRevE.92.023305}
  {\bibfield  {journal} {\bibinfo  {journal} {Phys. Rev. E}\ }\textbf {\bibinfo
  {volume} {92}},\ \bibinfo {pages} {023305} (\bibinfo {year}
  {2015})}\BibitemShut {NoStop}%
\bibitem [{\citenamefont {Macchi}\ \emph {et~al.}(2013)\citenamefont {Macchi},
  \citenamefont {Borghesi},\ and\ \citenamefont {Passoni}}]{Macchi2013rmp}%
  \BibitemOpen
  \bibfield  {author} {\bibinfo {author} {\bibfnamefont {A.}~\bibnamefont
  {Macchi}}, \bibinfo {author} {\bibfnamefont {M.}~\bibnamefont {Borghesi}},\
  and\ \bibinfo {author} {\bibfnamefont {M.}~\bibnamefont {Passoni}},\
  }\bibfield  {title} {\bibinfo {title} {Ion acceleration by superintense
  laser-plasma interaction},\ }\href
  {https://doi.org/10.1103/RevModPhys.85.751} {\bibfield  {journal} {\bibinfo
  {journal} {Rev. Mod. Phys.}\ }\textbf {\bibinfo {volume} {85}},\ \bibinfo
  {pages} {751} (\bibinfo {year} {2013})}\BibitemShut {NoStop}%
\bibitem [{\citenamefont {{Song}}\ \emph {et~al.}()\citenamefont {{Song}},
  \citenamefont {{Wang}},\ and\ \citenamefont {{Li}}}]{Song2021YUNIC}%
  \BibitemOpen
  \bibfield  {author} {\bibinfo {author} {\bibfnamefont {H.-H.}\ \bibnamefont
  {{Song}}}, \bibinfo {author} {\bibfnamefont {W.-M.}\ \bibnamefont {{Wang}}},\
  and\ \bibinfo {author} {\bibfnamefont {Y.-T.}\ \bibnamefont {{Li}}},\
  }\href@noop {} {\bibinfo {title} {{YUNIC}: A multi-dimensional
  particle-in-cell code for laser-plasma interaction}},\ \Eprint
  {https://arxiv.org/abs/2104.00642} {arXiv:2104.00642} \BibitemShut {NoStop}%
\bibitem [{\citenamefont {Song}\ \emph {et~al.}(2021)\citenamefont {Song},
  \citenamefont {Wang}, \citenamefont {Li}, \citenamefont {Li}, \citenamefont
  {Li}, \citenamefont {Sheng}, \citenamefont {Chen},\ and\ \citenamefont
  {Zhang}}]{Song2021njp}%
  \BibitemOpen
  \bibfield  {author} {\bibinfo {author} {\bibfnamefont {H.-H.}\ \bibnamefont
  {Song}}, \bibinfo {author} {\bibfnamefont {W.-M.}\ \bibnamefont {Wang}},
  \bibinfo {author} {\bibfnamefont {Y.-F.}\ \bibnamefont {Li}}, \bibinfo
  {author} {\bibfnamefont {B.-J.}\ \bibnamefont {Li}}, \bibinfo {author}
  {\bibfnamefont {Y.-T.}\ \bibnamefont {Li}}, \bibinfo {author} {\bibfnamefont
  {Z.-M.}\ \bibnamefont {Sheng}}, \bibinfo {author} {\bibfnamefont {L.-M.}\
  \bibnamefont {Chen}},\ and\ \bibinfo {author} {\bibfnamefont
  {J.}~\bibnamefont {Zhang}},\ }\bibfield  {title} {\bibinfo {title} {Spin and
  polarization effects on the nonlinear {B}reit{\textendash}{W}heeler pair
  production in laser-plasma interaction},\ }\href
  {https://doi.org/10.1088/1367-2630/ac0dec} {\bibfield  {journal} {\bibinfo
  {journal} {New J. Phys.}\ }\textbf {\bibinfo {volume} {23}},\ \bibinfo
  {pages} {075005} (\bibinfo {year} {2021})}\BibitemShut {NoStop}%
\bibitem [{\citenamefont {Li}\ \emph {et~al.}(2020{\natexlab{a}})\citenamefont
  {Li}, \citenamefont {Chen}, \citenamefont {Wang},\ and\ \citenamefont
  {Hu}}]{Li2020prl2}%
  \BibitemOpen
  \bibfield  {author} {\bibinfo {author} {\bibfnamefont {Y.-F.}\ \bibnamefont
  {Li}}, \bibinfo {author} {\bibfnamefont {Y.-Y.}\ \bibnamefont {Chen}},
  \bibinfo {author} {\bibfnamefont {W.-M.}\ \bibnamefont {Wang}},\ and\
  \bibinfo {author} {\bibfnamefont {H.-S.}\ \bibnamefont {Hu}},\ }\bibfield
  {title} {\bibinfo {title} {Production of highly polarized positron beams via
  helicity transfer from polarized electrons in a strong laser field},\ }\href
  {https://doi.org/10.1103/PhysRevLett.125.044802} {\bibfield  {journal}
  {\bibinfo  {journal} {Phys. Rev. Lett.}\ }\textbf {\bibinfo {volume} {125}},\
  \bibinfo {pages} {044802} (\bibinfo {year} {2020}{\natexlab{a}})}\BibitemShut
  {NoStop}%
\bibitem [{\citenamefont {Li}\ \emph {et~al.}(2020{\natexlab{b}})\citenamefont
  {Li}, \citenamefont {Shaisultanov}, \citenamefont {Chen}, \citenamefont
  {Wan}, \citenamefont {Hatsagortsyan}, \citenamefont {Keitel},\ and\
  \citenamefont {Li}}]{Li2020prl}%
  \BibitemOpen
  \bibfield  {author} {\bibinfo {author} {\bibfnamefont {Y.-F.}\ \bibnamefont
  {Li}}, \bibinfo {author} {\bibfnamefont {R.}~\bibnamefont {Shaisultanov}},
  \bibinfo {author} {\bibfnamefont {Y.-Y.}\ \bibnamefont {Chen}}, \bibinfo
  {author} {\bibfnamefont {F.}~\bibnamefont {Wan}}, \bibinfo {author}
  {\bibfnamefont {K.~Z.}\ \bibnamefont {Hatsagortsyan}}, \bibinfo {author}
  {\bibfnamefont {C.~H.}\ \bibnamefont {Keitel}},\ and\ \bibinfo {author}
  {\bibfnamefont {J.-X.}\ \bibnamefont {Li}},\ }\bibfield  {title} {\bibinfo
  {title} {Polarized ultrashort brilliant multi-{GeV} $\ensuremath{\gamma}$
  rays via single-shot laser-electron interaction},\ }\href
  {https://doi.org/10.1103/PhysRevLett.124.014801} {\bibfield  {journal}
  {\bibinfo  {journal} {Phys. Rev. Lett.}\ }\textbf {\bibinfo {volume} {124}},\
  \bibinfo {pages} {014801} (\bibinfo {year} {2020}{\natexlab{b}})}\BibitemShut
  {NoStop}%
\bibitem [{\citenamefont {Yokoya}(2011)}]{Cain}%
  \BibitemOpen
  \bibfield  {author} {\bibinfo {author} {\bibfnamefont {K.}~\bibnamefont
  {Yokoya}},\ }\href@noop {} {\emph {\bibinfo {title} {User's Manual of {CAIN}
  Version 2.42}}} (\bibinfo {year} {2011})\BibitemShut {NoStop}%
\bibitem [{mat()}]{material}%
  \BibitemOpen
  \href@noop {} {}\bibinfo {note} {See the Supplemental Material for detailed
  simulation methods and additional simulation results}\BibitemShut {NoStop}%
\bibitem [{\citenamefont {Esirkepov}\ \emph {et~al.}(2015)\citenamefont
  {Esirkepov}, \citenamefont {Bulanov}, \citenamefont {Koga}, \citenamefont
  {Kando}, \citenamefont {Kondo}, \citenamefont {Rosanov}, \citenamefont
  {Korn},\ and\ \citenamefont {Bulanov}}]{Esirkepov2015pla}%
  \BibitemOpen
  \bibfield  {author} {\bibinfo {author} {\bibfnamefont {T.~Z.}\ \bibnamefont
  {Esirkepov}}, \bibinfo {author} {\bibfnamefont {S.~S.}\ \bibnamefont
  {Bulanov}}, \bibinfo {author} {\bibfnamefont {J.~K.}\ \bibnamefont {Koga}},
  \bibinfo {author} {\bibfnamefont {M.}~\bibnamefont {Kando}}, \bibinfo
  {author} {\bibfnamefont {K.}~\bibnamefont {Kondo}}, \bibinfo {author}
  {\bibfnamefont {N.~N.}\ \bibnamefont {Rosanov}}, \bibinfo {author}
  {\bibfnamefont {G.}~\bibnamefont {Korn}},\ and\ \bibinfo {author}
  {\bibfnamefont {S.~V.}\ \bibnamefont {Bulanov}},\ }\bibfield  {title}
  {\bibinfo {title} {Attractors and chaos of electron dynamics in
  electromagnetic standing waves},\ }\href
  {https://doi.org/https://doi.org/10.1016/j.physleta.2015.06.017} {\bibfield
  {journal} {\bibinfo  {journal} {Phys. Lett. A}\ }\textbf {\bibinfo {volume}
  {379}},\ \bibinfo {pages} {2044 } (\bibinfo {year} {2015})}\BibitemShut
  {NoStop}%
\bibitem [{\citenamefont {Song}\ \emph {et~al.}(2019)\citenamefont {Song},
  \citenamefont {Wang}, \citenamefont {Li}, \citenamefont {Li},\ and\
  \citenamefont {Li}}]{Song2019pra}%
  \BibitemOpen
  \bibfield  {author} {\bibinfo {author} {\bibfnamefont {H.-H.}\ \bibnamefont
  {Song}}, \bibinfo {author} {\bibfnamefont {W.-M.}\ \bibnamefont {Wang}},
  \bibinfo {author} {\bibfnamefont {J.-X.}\ \bibnamefont {Li}}, \bibinfo
  {author} {\bibfnamefont {Y.-F.}\ \bibnamefont {Li}},\ and\ \bibinfo {author}
  {\bibfnamefont {Y.-T.}\ \bibnamefont {Li}},\ }\bibfield  {title} {\bibinfo
  {title} {Spin-polarization effects of an ultrarelativistic electron beam in
  an ultraintense two-color laser pulse},\ }\href
  {https://doi.org/10.1103/PhysRevA.100.033407} {\bibfield  {journal} {\bibinfo
   {journal} {Phys. Rev. A}\ }\textbf {\bibinfo {volume} {100}},\ \bibinfo
  {pages} {033407} (\bibinfo {year} {2019})}\BibitemShut {NoStop}%
\bibitem [{\citenamefont {Brodin}\ and\ \citenamefont
  {Marklund}(2007)}]{Brodin2007njp}%
  \BibitemOpen
  \bibfield  {author} {\bibinfo {author} {\bibfnamefont {G.}~\bibnamefont
  {Brodin}}\ and\ \bibinfo {author} {\bibfnamefont {M.}~\bibnamefont
  {Marklund}},\ }\bibfield  {title} {\bibinfo {title} {Spin
  magnetohydrodynamics},\ }\href {https://doi.org/10.1088/1367-2630/9/8/277}
  {\bibfield  {journal} {\bibinfo  {journal} {New J. Phys.}\ }\textbf {\bibinfo
  {volume} {9}},\ \bibinfo {pages} {277} (\bibinfo {year} {2007})}\BibitemShut
  {NoStop}%
\end{thebibliography}

\clearpage


\section*{Supplementary material}

# Dense Polarized Positrons from Laser-Irradiated Foil Targets in the QED Regime (Supplemental Material)

Huai-Hang Song,[1, 3] Wei-Min Wang,[2, 4, *] and Yu-Tong Li[1, 3, 5, †]

[1]*Beijing National Laboratory for Condensed Matter Physics,*
*Institute of Physics, Chinese Academy of Sciences, Beijing 100190, China*
[2]*Department of Physics and Beijing Key Laboratory of Opto-electronic Functional Materials and Micro–nano Devices,*
*Renmin University of China, Beijing 100872, China*
[3]*School of Physical Sciences, University of Chinese Academy of Sciences, Beijing 100049, China*
[4]*Collaborative Innovation Center of IFSA, Shanghai Jiao Tong University, Shanghai 200240, China*
[5]*Songshan Lake Materials Laboratory, Dongguan, Guangdong 523808, China*
(Dated: December 13, 2021)

## I. QED PROBABILITIES AND MONTE-CARLO METHODS

In this work, the spin- and polarization-resolved probabilities of photon emission and $e^-e^+$ pair production derived by Baier and Katkov are employed, which are written as [1–3]

$$\frac{d^2W_\gamma}{du dt} = \frac{C_{\rm rad}}{4}\left\{\frac{u^2-2u+2}{1-u}K_{2/3}(y_1) - {\rm Int}K_{1/3}(y_1) - uK_{1/3}(y_1)(\bm{S}_i\cdot\bm{e}_2)\right.$$
$$-\frac{u}{1-u}K_{1/3}(y_1)(\bm{S}_f\cdot\bm{e}_2) + \left[2K_{2/3}(y_1) - {\rm Int}K_{1/3}(y_1)\right](\bm{S}_i\cdot\bm{S}_f)$$
$$+\frac{u^2}{1-u}\left[K_{2/3}(y_1) - {\rm Int}K_{1/3}(y_1)\right](\bm{S}_i\cdot\bm{e}_v)(\bm{S}_f\cdot\bm{e}_v)$$
$$+\frac{u}{1-u}K_{1/3}(y_1)(\bm{S}_i\cdot\bm{e}_1)\xi_1 + \left[\frac{2u-u^2}{1-u}K_{2/3}(y_1) - u{\rm Int}K_{1/3}(y_1)\right](\bm{S}_i\cdot\bm{e}_v)\xi_2$$
$$\left.+\left[K_{2/3}(y_1) - \frac{u}{1-u}K_{1/3}(y_1)(\bm{S}_i\cdot\bm{e}_2)\right]\xi_3\right\}, \tag{S1}$$

$$\frac{d^2W_\pm}{d\varepsilon_\pm dt} = \frac{C_{\rm pairs}}{2}\left\{\frac{\varepsilon_+^2+\varepsilon_-^2}{\varepsilon_+\varepsilon_-}K_{2/3}(y_2) + {\rm Int}K_{1/3}(y_2) - \xi_3'K_{2/3}(y_2)\right.$$
$$-\xi_1'\frac{\varepsilon_\gamma}{\varepsilon_\mp}K_{1/3}(y_2)(\bm{S}_\pm\cdot\bm{e}_1') + \xi_2'\left[\frac{\varepsilon_\gamma}{\varepsilon_\pm}{\rm Int}K_{1/3}(y_2) + \frac{\varepsilon_\pm^2-\varepsilon_\mp^2}{\varepsilon_+\varepsilon_-}K_{2/3}(y_2)\right](\bm{S}_\pm\cdot\bm{e}_v)$$
$$\left.-(\frac{\varepsilon_\gamma}{\varepsilon_\pm} - \xi_3'\frac{\varepsilon_\gamma}{\varepsilon_\mp})K_{1/3}(y_2)(\bm{S}_\pm\cdot\bm{e}_2')\right\}, \tag{S2}$$

where $K_\nu(y)$ is the modified Bessel function of the second kind with a noninteger factor $\nu$, ${\rm Int}K_{1/3}(y) \equiv \int_y^\infty K_{1/3}(x)dx$, $y_1 = 2u/[3(1-u)\chi_e]$, $y_2 = 2\varepsilon_\gamma^2/(3\chi_\gamma\varepsilon_+\varepsilon_-)$, $u = \varepsilon_\gamma/\varepsilon_e$, $C_{\rm rad} = (\alpha m_e^2 c^4)/(\sqrt{3}\pi\hbar\varepsilon_e)$, $C_{\rm pairs} = (\alpha m_e^2 c^4)/(\sqrt{3}\pi\hbar\varepsilon_\gamma^2)$, and $\alpha \approx 1/137$ is the fine structure constant. Particle energies $\varepsilon_e$, $\varepsilon_\gamma$, $\varepsilon_+$, and $\varepsilon_-$ are those of emitting leptons (electrons or positrons), emitted $\gamma$ photons, and the created positrons and electrons, respectively. $\bm{S}_i$ and $\bm{S}_f$ are the spin vectors of leptons before and after the radiation. $\bm{S}_-$ and $\bm{S}_+$ are the spin vectors of the newly created electrons and positrons. When Eqs. (S1) and (S2) are applied to the electron (positron), $\bm{e}_v$ represents the unit vector along the electron (positron) velocity, $\bm{e}_1$ and $\bm{e}_1'$ represent the unit vectors along the transverse electron (positron) acceleration, $\bm{e}_2 = \bm{e}_v\times\bm{e}_1$ and $\bm{e}_2' = \bm{e}_v\times\bm{e}_1'$, respectively. The Stokes parameter $\bm{\xi}$ is defined with the orthonormal basis $(\bm{e}_1, \bm{e}_2, \bm{e}_v)$, while $\bm{\xi}'$ is defined with $(\bm{e}_1', \bm{e}_2', \bm{e}_v)$. Based on the locally constant-field approximation, the QED model above holds for the ultraintense laser field of the normalized intensity $a_0 \gg 1$ [4].

The spin and polarization relevant Monte-Carlo algorithms in our QED-PIC code YUNIC [5, 6] are basically based on the single-particle code CAIN [7], in which the mean axes of leptons and photons are selected to be their quantization axes. After a photon emission, the lepton spin flips with respect to $\bm{S}_{\rm R} = \pm\bm{S}_{\rm R}^*/|\bm{S}_{\rm R}^*|$, where

---
* weiminwang1@ruc.edu.cn
† ytli@iphy.ac.cn

$\boldsymbol{S}_{\mathrm{R}}^* = \left[2K_{2/3}(y_1) - \mathrm{Int}K_{1/3}(y_1)\right]\boldsymbol{S}_i - \frac{u}{1-u}K_{1/3}(y_1)\boldsymbol{e}_2 + \frac{u^2}{1-u}\left[K_{2/3}(y_1) - \mathrm{Int}K_{1/3}(y_1)\right](\boldsymbol{S}_i \cdot \boldsymbol{e}_v)\boldsymbol{e}_v$. The Stokes parameters of the emitted photon at birth are chosen as $\boldsymbol{\xi} = \pm \boldsymbol{\xi}^*/|\boldsymbol{\xi}^*|$, where $\boldsymbol{\xi}^* = (\xi_1^*, \xi_2^*, \xi_3^*)$, $\xi_1^* = \frac{u}{1-u}K_{1/3}(y_1)(\boldsymbol{S}_i \cdot \boldsymbol{e}_1)$, $\xi_2^* = \left[\frac{2u-u^2}{1-u}K_{2/3}(y_1) - u\mathrm{Int}K_{1/3}(y_1)\right](\boldsymbol{S}_i \cdot \boldsymbol{e}_v)$, and $\xi_3^* = K_{2/3}(y_1) - \frac{u}{1-u}K_{1/3}(y_1)(\boldsymbol{S}_i \cdot \boldsymbol{e}_2)$. The spin vector of created pairs at birth is parallel or antiparallel to $\boldsymbol{S}_\pm^*/|\boldsymbol{S}_\pm^*|$, where $\boldsymbol{S}_\pm^* = -\xi_1'(\varepsilon_\gamma/\varepsilon_\mp)K_{1/3}(y_2)\boldsymbol{e}_1' - (\varepsilon_\gamma/\varepsilon_\pm - \xi_3'\varepsilon_\gamma/\varepsilon_\mp)K_{1/3}(y_2)\boldsymbol{e}_2' + \xi_2'\left[(\varepsilon_\gamma/\varepsilon_\pm)\mathrm{Int}K_{1/3}(y_2) + (\varepsilon_\pm^2 - \varepsilon_\mp^2)/(\varepsilon_+\varepsilon_-)K_{2/3}(y_2)\right]\boldsymbol{e}_v$.

The spin vector of leptons $\boldsymbol{S}_{\mathrm{NR}}$ without photon emissions and the Stokes parameters of photons $\boldsymbol{\xi}_{\mathrm{ND}}$ before decaying into $e^-e^+$ pairs must be treated carefully due to selection effects as pointed in CAIN [7]. This is because we have chosen the mean axes as the quantization axes, indicating that every particle in the simulation actually represents an assemble of particles. Taking the radiative polarization process as an example. The average polarization of non-emitting leptons would change since the the photon emission probability $d^2W_\gamma/(dudt)$ relies on the lepton spin vector $\boldsymbol{S}_i$ [see the third term $-uK_{1/3}(y_1)(\boldsymbol{S}_i \cdot \boldsymbol{e}_2)$ in Eq. (S1)]. It means that the non-emitting leptons are more prone to be $\boldsymbol{S}_i \cdot \boldsymbol{e}_2 > 0$ due to the smaller emission probability $d^2W_\gamma/(dudt)$ compared with those of $\boldsymbol{S}_i \cdot \boldsymbol{e}_2 < 0$, which causes their mean spin vectors to deviate from the initial ones. Therefore, for non-emitting leptons, their spin vectors should also be updated according to another quantization axis $\boldsymbol{S}_{\mathrm{NR}} = \pm \boldsymbol{S}_{\mathrm{NR}}^*/|\boldsymbol{S}_{\mathrm{NR}}^*|$, where $\boldsymbol{S}_{\mathrm{NR}}^* = \boldsymbol{S}_i\left\{1 - C_{\mathrm{rad}}\Delta t \int_0^1 \left[\frac{u^2-2u+2}{1-u}K_{2/3}(y_1) - \mathrm{Int}K_{1/3}(y_1)\right]du\right\} + \boldsymbol{e}_2 C_{\mathrm{rad}}\Delta t \int_0^1 uK_{1/3}(y_1)du$ [3, 7]. The similar selection effect should also be applied to non-decaying photons because the third term $-\xi_3' K_{2/3}(y_2)$ in Eq. (S2) also plays a selection role in the photon polarization. The quantization axes of non-decaying photons are $\boldsymbol{\xi}_{\mathrm{ND}} = \pm \boldsymbol{\xi}_{\mathrm{ND}}^*/|\boldsymbol{\xi}_{\mathrm{ND}}^*|$, where $\boldsymbol{\xi}_{\mathrm{ND}}^* = \boldsymbol{\xi}\left\{1 - C_{\mathrm{pairs}}\Delta t \int_0^{\varepsilon_\gamma}\left[\frac{\varepsilon_+^2+\varepsilon_-^2}{\varepsilon_+\varepsilon_-}K_{2/3}(y_2) + \mathrm{Int}K_{1/3}(y_2)\right]d\varepsilon_+\right\} + \boldsymbol{e}^* C_{\mathrm{pairs}}\Delta t \int_0^{\varepsilon_\gamma} K_{2/3}(y_2)d\varepsilon_+$ and $\boldsymbol{e}^* = (0,0,1)$. Note that the selection effect of non-emitting leptons are extensively considered recently, while that of non-decaying photons has not attracted much attention.

To illustrate the importance of the selection effect, we consider 10 GeV electrons or photons are injected into the perpendicular static magnetic field of a strength $B_0 = 1.07 \times 10^6$ T. Three Monte-Carlo methods based on different quantization axes are employed: (I) the mean axis, as described above; (II) the fixed axis, i.e., $\pm \boldsymbol{e}_2$ axis for leptons and $(0,0,\pm 1)$ axis for photons, which does not require additional calculations for non-emitting leptons or non-decaying photons; (III) the mean axis but excluding the selection effects. The time evolution of the average transverse polarization $\overline{S}_t$ of primary electrons and the average linear polarization $\overline{\xi}_3$ of primary photons are showed in Figs. S2(a) and S2(b), respectively. We can see that methods I and II are in good agreement, while method III cannot give the correct results.

Although method II can give the same results as method I in the case of a static magnetic field above, it cannot preserve the completely 3D polarization dynamics. Therefore, our employed method I is more suitable for the complex electromagnetic environments in laser-plasma interactions.

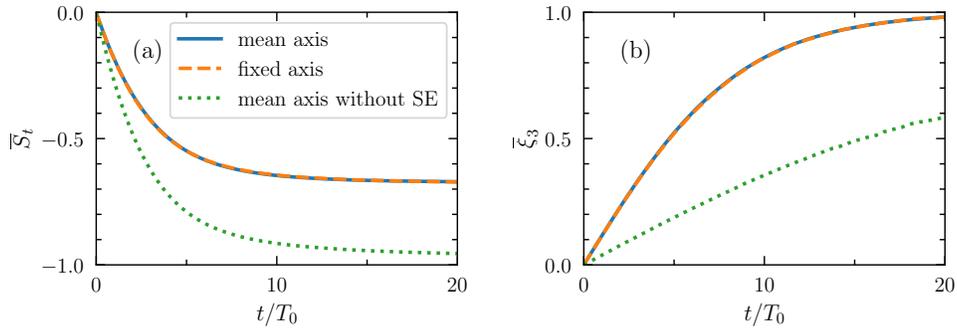

FIG. S1. Time evolution of (a) transverse polarization of primary electrons and (b) linear polarization of primary photons (the annihilated primary photons and secondary photons are not counted) by choosing the mean axis (blue-solid), fixed axis (orange-dashed), and mean axis but without selection effects (SE) (green-dotted), respectively, where $T_0 = 3.33$ fs.

## II. POLARIZATIONS OF LEPTONS AND PHOTONS

Here, we utilize the simplified models specific to this work to analyze the involved polarization processes. In our considered configuration, the linear laser polarization along the $y$ direction leads to that leptons move almost in the laser polarization plane ($x$-$y$ plane). Note that $\boldsymbol{S}$ is defined in the lepton rest frame, its average spin component $\overline{S}_z$ can always characterize the transverse polarization since positrons move completely (mainly) in the $x$-$y$ plane for 1D

(later 3D) geometry. The electromagnetic fields in the rest frame of an ultrarelativistic lepton is

$$\boldsymbol{E}' = \gamma(\boldsymbol{E} + \boldsymbol{\beta} \times \boldsymbol{B}) - \frac{\gamma^2}{\gamma+1}\boldsymbol{\beta}(\boldsymbol{\beta}\cdot\boldsymbol{E}) \approx \gamma F_0(\beta_y, -\beta_x, 0), \tag{S3}$$

$$\boldsymbol{B}' = \gamma(\boldsymbol{B} - \boldsymbol{\beta} \times \boldsymbol{E}) - \frac{\gamma^2}{\gamma+1}\boldsymbol{\beta}(\boldsymbol{\beta}\cdot\boldsymbol{B}) \approx \gamma F_0(0, 0, 1), \tag{S4}$$

where $\gamma$ is the lepton relativistic factor, $\boldsymbol{\beta} \approx (\beta_x, \beta_y, 0)$ is the unit vector along the lepton velocity, and $F_0 = B_z - \beta_x E_y$. As expected, the electric field $\boldsymbol{E}'$ and magnetic field $\boldsymbol{B}'$ of the same magnitude are orthogonal to $\boldsymbol{\beta}$. In the magnetic field dominated regime where the photon emission and the pair creation mainly happen, the rest-frame magnetic field direction $\boldsymbol{\zeta} \equiv \boldsymbol{B}'/|\boldsymbol{B}'|$ is approximated along the laser magnetic field direction of the laboratory frame, i.e. $\boldsymbol{\zeta} \approx (0, 0, B_z/|B_z|)$. For convenience of expression, the spin vector of electrons is defined with respect to $-\boldsymbol{\zeta}$, while that of positrons is with respect to $\boldsymbol{\zeta}$; in other words, the same polarization values for electrons and positrons means that their spins are directed oppositely here. In addition, electrons and positrons cannot acquire the longitudinal polarization due to an initially unpolarized target that we take, and the emitted photons are only linearly polarized with $\xi_3 \neq 0$. We no longer distinguish between $\xi_3$ and $\xi_3'$ due to $\xi_3 \approx \xi_3'$ in the linearly polarized laser fields [6]. Equations (S1) and (S2) can therefore be simplified to

$$\frac{d^2 W_\gamma}{dudt} = \frac{C_{\rm rad}}{4} \left\{ \frac{u^2-2u+2}{1-u}K_{2/3}(y_1) - {\rm Int}K_{1/3}(y_1) + uK_{1/3}(y_1)S_i \right. $$
$$+ \frac{u}{1-u}K_{1/3}(y_1)S_f + \left[2K_{2/3}(y_1) - {\rm Int}K_{1/3}(y_1)\right]S_i S_f $$
$$\left. + \left[K_{2/3}(y_1) + \frac{u}{1-u}K_{1/3}(y_1)S_i\right]\xi_3 \right\}, \tag{S5}$$

$$\frac{d^2 W_\pm}{d\varepsilon_\pm dt} = \frac{C_{\rm pairs}}{2} \left\{ \frac{\varepsilon_+^2 + \varepsilon_-^2}{\varepsilon_+\varepsilon_-}K_{2/3}(y_2) + {\rm Int}K_{1/3}(y_2) - \xi_3 K_{2/3}(y_2) \right.$$
$$\left. + \left(\frac{\varepsilon_\gamma}{\varepsilon_\pm} - \xi_3 \frac{\varepsilon_\gamma}{\varepsilon_\mp}\right) K_{1/3}(y_2)S_\pm \right\}. \tag{S6}$$

Accordingly, we can obtain the averaged polarization $\overline{S}_f$, $\overline{\xi}_3$ and $\overline{S}_\pm$, respectively.

1. The average polarization of the positron after a photon emission:

$$\overline{S}_f = \frac{\frac{u}{1-u}K_{1/3}(y_1) + \left[2K_{2/3}(y_1) - {\rm Int}K_{1/3}(y_1)\right]S_i}{\frac{u^2-2u+2}{1-u}K_{2/3}(y_1) - {\rm Int}K_{1/3}(y_1) + uK_{1/3}(y_1)S_i}. \tag{S7}$$

2. The average polarization of the emitted photon:

$$\overline{\xi}_3 = \frac{K_{2/3}(y_1) + \frac{u}{1-u}K_{1/3}(y_1)S_i}{\frac{u^2-2u+2}{1-u}K_{2/3}(y_1) - {\rm Int}K_{1/3}(y_1) + uK_{1/3}(y_1)S_i}. \tag{S8}$$

3. The average polarization of the newly produced positron:

$$\overline{S}_\pm = \frac{\left(\frac{\varepsilon_\gamma}{\varepsilon_\pm} - \xi_3 \frac{\varepsilon_\gamma}{\varepsilon_\mp}\right) K_{1/3}(y_2)}{\left(\frac{\varepsilon_+^2+\varepsilon_-^2}{\varepsilon_+\varepsilon_-} - \xi_3\right) K_{2/3}(y_2) + {\rm Int}K_{1/3}(y_2)}. \tag{S9}$$

These polarizations exhibit distinct characteristics between low- and high-energy regimes. For the photon emission as shown in Figs. S2(a) and S2(b), when emitting a low-energy photon of $\varepsilon_\gamma/\varepsilon_e \ll 1$, the lepton nearly keeps its polarization unchanged, and the average photon polarization always has a positive value of around 0.5, insensitive to the lepton polarization. While for a high-energy emitted photon $\varepsilon_\gamma/\varepsilon_e \to 1$, the lepton spin tends to flip along $\boldsymbol{\zeta}$, which is parallel (antiparallel) to $\boldsymbol{B}'$ for positrons (electrons); meanwhile, the polarization of emitted photon is highly dependent of the lepton polarization, with $\overline{\xi}_3 \approx S_i$. For the $e^-e^+$ pair production as shown in Fig. S2(c), the spin of low-energy lepton is more likely to be parallel to $\boldsymbol{\zeta}$, but the high-energy case is determined by the photon polarization. Above asymptotic relationships under low- and high-energy limits can be summarized as

$$\overline{S}_f = \begin{cases} S_i, & \varepsilon_\gamma/\varepsilon_e \to 0 \\ 1, & \varepsilon_\gamma/\varepsilon_e \to 1 \end{cases}, \quad \overline{\xi}_3 = \begin{cases} 0.5, & \varepsilon_\gamma/\varepsilon_e \to 0 \\ S_i, & \varepsilon_\gamma/\varepsilon_e \to 1 \end{cases}, \quad \overline{S}_\pm = \begin{cases} 1, & \varepsilon_\pm/\varepsilon_\gamma \to 0 \\ -\xi_3, & \varepsilon_\pm/\varepsilon_\gamma \to 1 \end{cases}$$

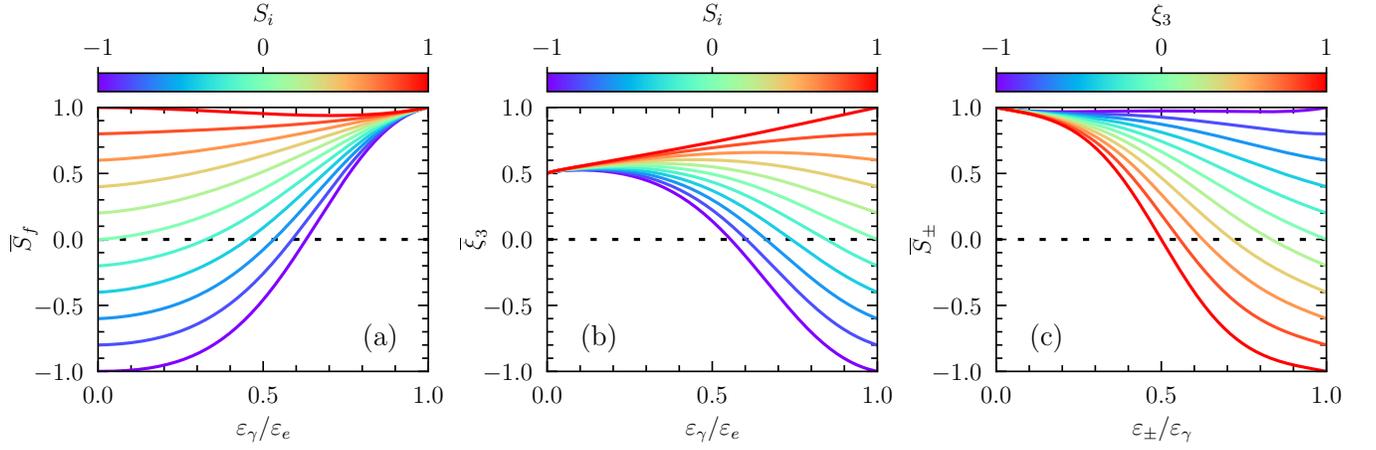

FIG. S2. (a) The average lepton polarization $\overline{S}_f$ after radiation and (b) the average emitted photon polarization $\overline{\xi}_3$ versus the energy fraction $\varepsilon_\gamma/\varepsilon_e$ for various lepton polarizations $S_i$ before radiation with the QED parameter $\chi_e = 2$. The average polarization $\overline{S}_\pm$ of generated leptons via the $e^-e^+$ pair production versus the energy fraction $\varepsilon_\pm/\varepsilon_\gamma$ for various photon polarizations $\xi_3$ with the QED parameter $\chi_\gamma = 2$.

## III. ADDITIONAL SIMULATION RESULTS

### A. Photon polarization

For an initially unpolarized foil target, the emitted $\gamma$ photons have a positive polarization $\overline{\xi}_3 > 0$, which is decreased with the photon energy $\varepsilon_\gamma$ as shown in Fig. S3. This is consistent with the theoretical plot of Fig. S2(b). It means that for high-energy photons that are responsible for the $e^-e^+$ pair production, they are weakly polarized. Therefore, according to Fig. S2(c), the generated positrons have a positive polarization $\overline{S}_z > 0$ in the positive laser cycle $B_z > 0$, while a negative value $\overline{S}_z < 0$ in the negative laser cycle $B_z < 0$, i.e., their spin vectors are dominantly parallel to laser magnetic field $\boldsymbol{B}$. According to the third term $-\xi_3 K_{2/3}(y_2)$ of Eq. (S6), the positive $\overline{\xi}_3$ would decrease the $e^-e^+$ pair production probability, leading to a 7% positron yield decrease observed in our simulations compared to that excluding the spin and polarization effects.

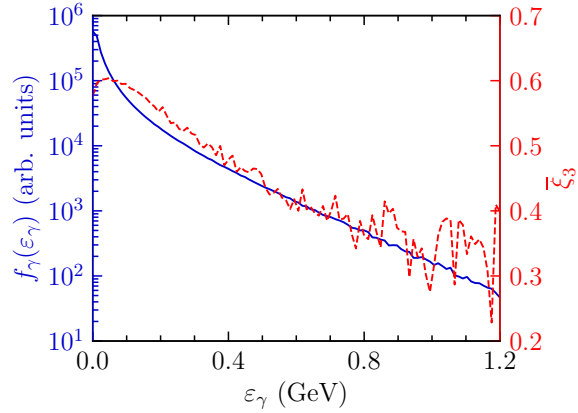

FIG. S3. The photon polarization $\overline{\xi}_3$ versus the photon energy $\varepsilon_\gamma$ and photon energy spectrum $f_\gamma(\varepsilon_\gamma)$. All parameters are the same as Fig. (1).

### B. Electron polarization

Figures S4(a) and S4(b) show the electron polarization (including initial target electrons and later generated electrons). For the generated electrons of $e^-e^+$ pairs, they can also be polarized like positrons, with $\overline{S}_z > 0$ at $\theta_y < 0$ and

$\overline{S}_z < 0$ at $\theta_y > 0$. However, due to the existence of substantial unpolarized target electrons, the global polarization degree is weakened to some extent. From the blue line of Fig. S4(b), the electron polarization can arrive $\sim 20\%$ at $|\theta_y| > 20°$, smaller that 30% of positrons. With some certain deflection angles and energies, the electron polarization can also reach around 60%.

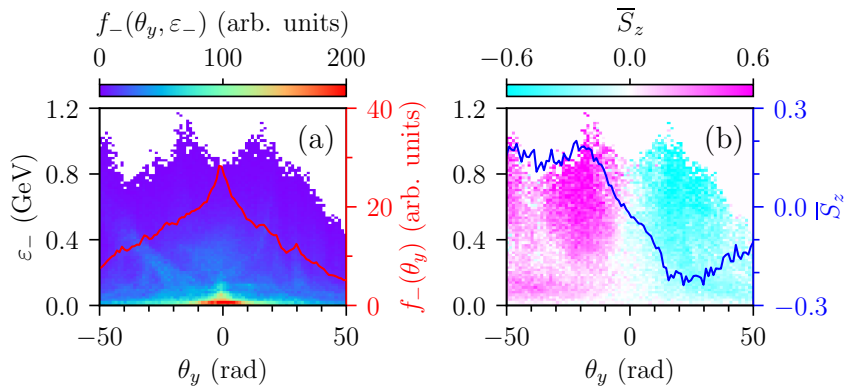

FIG. S4. (a) The number distribution $f_-(\theta_y, \varepsilon_-)$ and (b) polarization $\overline{S}_z$ of electrons versus the deflection angle $\theta_y$ and electron energy $\varepsilon_-$ at the end of the interaction $t = 28T_0$. All parameters are the same as those in Fig. (1).

### C. Polarization of positrons born in the positive half-cycle

For comparison, we also present the polarization properties of positrons born in a positive laser cycle in Figs. S5(a) and S5(b) [next to the elliptical zone of Fig. 2(a) we previously focused]. As born in the positive laser cycle of $B_z > 0$, positrons at birth would be polarized along the $\boldsymbol{B}$ direction to achieve $\overline{S}_z > 0$, as shown in Fig. S5(a), which is opposite to the negative cycle case of $\overline{S}_z < 0$ in Fig. 3(a). Therefore, the initial positron polarization between positive and negative cycles vanishes. Under the radiation reaction and radiative polarization, positrons deflected along the $\pm y$ directions are polarized along the $\mp z$ directions, which is consistent with the negative cycle case in Fig. 3(b), thus leading to the global positron polarization.

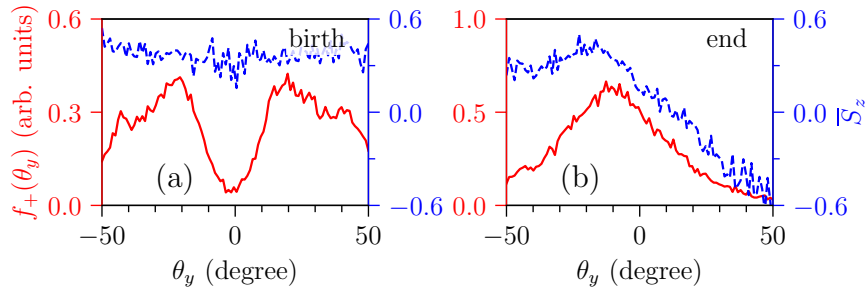

FIG. S5. For positrons born in a positive laser cycle. The number distribution $f_+(\theta_y)$ and polarization $\overline{S}_z$ versus the deflection angle $\theta_y$ (a) at birth from $t = 21.4T_0$ to $22T_0$ and (b) at the end of the interaction $t = 28T_0$.

### D. Small-scale preplasma

In the case of a small scale-length of preplasmas, the laser pulse cannot quickly trigger the QED cascades near the target surface. We show in Fig. S6 that positrons are generated away from the target surface for $L = 0.4$ $\mu$m, rather than near the target surface in the case of $L = 1.5$ $\mu$m in Fig. 2(a). Therefore, after being generated, positrons would experience multicycle laser fields before escaping, consequently the polarization mechanism breaks, leading to a weaker polarization.

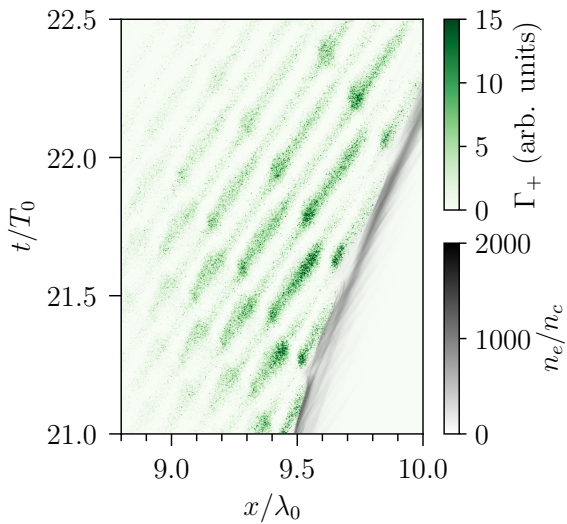

FIG. S6. The spatiotemporal evolution of electron density $n_e$ and positron production rate $\Gamma_+$ in the case of a small scale-length $L = 0.4$ $\mu$m. Other parameters are the same as Fig. 1.

### E. Influence of target density

Here, we investigate the influence of target density (i.e., $n_0 = 300n_c$, $530n_c$ and $700n_c$) on the positron yield $N_+$ and the positron polarization $\overline{S}_z$, as shown in Fig. S7(a) and S7(b), respectively. For the relatively lower-density target of $n_0 = 300n_c$, the positron yield is much less than the other two higher-density cases. Meanwhile, its optimal laser strength for the positron polarization is around $a_0 = 1500$, less than $a_0 = 1700$ of $n_0 = 530n_c$. This is because the light pressure is more likely to destroy the lower-density target with the laser strength increasing.

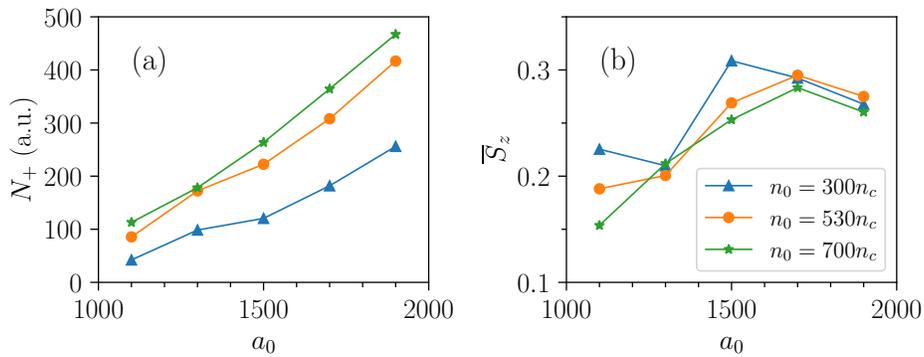

FIG. S7. The number $N_+$ and polarization $\overline{S}_z$ of positrons at $\theta_y > 10°$ versus the normalized laser peak strength $a_0$ under three target densities $n_0 = 300n_c$, $530n_c$, and $700n_c$, respectively. Other parameters are the same as Fig. (1).

### F. 3D simulation parameters and obliquely incident laser pulse at 30°

In Fig. 5, the 3D simulation domain is $L_x \times L_y \times L_z = 20\lambda_0 \times 12\lambda_0 \times 12\lambda_0$, resolved by $640 \times 384 \times 384$ cells, and filled with 25 electrons and 16 $C^{6+}$ ions per cell. Other laser and target parameters are the same as Fig. 1.

We have also performed an additional 3D simulation to investigate the positron polarization in the case of the obliquely p-polarized incident laser pulse at 30°, since in real experiments the oblique incidence is often employed to avoid the damage of reflected lights to optical devices. From Fig. S8, we can see that the obvious deflection-angle-dependent polarization is still observed. There are still 22 nC positrons of a polarization degree above 30%, only slightly smaller that 30 nC of normal incidence case. Thus, the proposed scheme is easily conducted in experiments


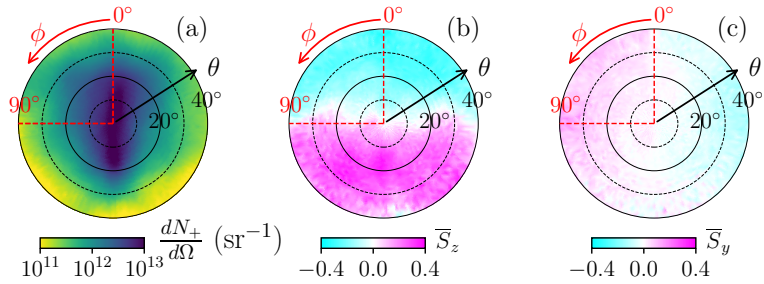

FIG. S8. Angular distributions of (a) positron number, as well as two positron polarization components (b) $\overline{S}_z$ and (c) $\overline{S}_y$ versus the polar angle $\theta$ and azimuthal angle $\phi$, where the laser pulse polarized along $\phi = 0, 180°$ propagates along $\theta = 0$. Other parameters are the same as those in Fig. 5 expect that the laser incident angle is $30°$.

with future 100-PW-class laser facilities.

[1] V. N. Baier, V. Katkov, and V. M. Strakhovenko, *Electromagnetic Processes at High Energies in Oriented Single Crystals* (World Scientific, Singapore, 1998).
[2] Y.-F. Li, R. Shaisultanov, Y.-Y. Chen, F. Wan, K. Z. Hatsagortsyan, C. H. Keitel, and J.-X. Li, Polarized ultrashort brilliant multi-GeV $\gamma$ rays via single-shot laser-electron interaction, Phys. Rev. Lett. **124**, 014801 (2020).
[3] Y.-F. Li, Y.-Y. Chen, W.-M. Wang, and H.-S. Hu, Production of highly polarized positron beams via helicity transfer from polarized electrons in a strong laser field, Phys. Rev. Lett. **125**, 044802 (2020).
[4] V. I. Ritus, Quantum effects of the interaction of elementary particles with an intense electromagnetic field, J. Sov. Laser Res. **6**, 497 (1985).
[5] H.-H. Song, W.-M. Wang, and Y.-T. Li, YUNIC: A multi-dimensional particle-in-cell code for laser-plasma interaction, arXiv:2104.00642.
[6] H.-H. Song, W.-M. Wang, Y.-F. Li, B.-J. Li, Y.-T. Li, Z.-M. Sheng, L.-M. Chen, and J. Zhang, Spin and polarization effects on the nonlinear Breit–Wheeler pair production in laser-plasma interaction, New J. Phys. **23**, 075005 (2021).
[7] K. Yokoya, *User's Manual of CAIN Version 2.42* (2011).



\end{document}